\documentclass[12pt]{article}
\usepackage{epsfig,amsmath,graphics}
\usepackage{graphics,graphicx}
\usepackage{tabularx}
\usepackage{amsfonts}
\usepackage{amsmath}
\usepackage{color}
\usepackage{epstopdf}
\usepackage[round, authoryear, comma, sort&compress]{natbib}
\newcommand{\beq}{\begin{equation}}
\newcommand{\eeq}{\end{equation}}
\newcommand{\lb}{\label}
\newcommand{\beqar}{\begin{eqnarray}}
\newcommand{\eeqar}{\end{eqnarray}}
\newcommand{\barr}{\begin{array}}
\newcommand{\earr}{\end{array}}

\renewcommand{\d}{\textrm{d}}

\def\XXint#1#2#3{{\setbox0=\hbox{$#1{#2#3}{\int}$}
     \vcenter{\hbox{$#2#3$}}\kern-.5\wd0}}

\def\bB{\mbox{\bf{B}}}

\def\bC{\mbox{\bf{C}}}

\def\bD{\mbox{\bf{D}}}

\def\bE{\mbox{\bf{E}}}

\def\bF{\mbox{\bf{F}}}
\def\bff{\mbox{\bf{f}}}

\def\bI{\mbox{\bf{I}}}

\def\bn{\mbox{\bf{n}}}
\def\b0{\mbox{\bf{0}}}

\def\bP{\mbox{\bf{P}}}

\def\bt{\mbox{\bf{t}}}

\def\bX{\mbox{\bf{X}}}
\def\bx{\mbox{\bf{x}}}

\def\bchi{\mbox{\boldmath${\chi}$}}

\def\bsigma{\mbox{\boldmath${\sigma}$}}

\def\tr{{\rm tr}}

\def\divv{{\rm div}}
\def\Div{{\rm Div}}

\def\curl{{\rm curl}}
\def\Curl{{\rm Curl}}
\def\Grad{{\rm Grad}}
\def\grad{{\rm grad}}

\def\diag{{\rm diag}}

\def\tlzp{\mbox{${\tilde\lambda_z}^{(p)}$}}
\def\tlza{\mbox{${\tilde\lambda_z}^{(a)}$}}

\def\lz{\mbox{${\lambda_z}$}}

\def\rmT{\mathrm{T}}
\def\rmc{\mathrm{c}}

\def\salto#1#2{
[\mbox{\hspace{-#1em}}[#2]\mbox{\hspace{-#1em}}]}

\begin{document}

\title{Analysis of multilayer electro-active tubes under different constraints}
\author{Eliana Bortot$^a$
\\
\\
\small{\sl{$^a$Structural Engineering Department}},
\\
\small{\sl{University of California San Diego, 9500 Gilman Drive, Mail Code 0085, CA 92093 San Diego USA.}}
\\
\small{\sl{Email: ebortot@eng.ucsd.edu}} \\
}

\maketitle

\begin{abstract}
	Dielectric elastomers are an emerging class of highly deformable electro-active materials employed for electromechanical transduction technology.
	For practical applications, the design of such transducers requires a model accounting for insulation of the active membrane, non-perfectly compliant behaviour of the electrodes or interaction of the transducer with a soft actuated body. To this end, a three-layer model, in which the active membrane is embedded between two soft passive layers, can be formulated.
	In this paper, the theory of nonlinear electro-elasticity for heterogeneous soft dielectrics is used to investigate the electromechanical response of multilayer electro-active tubes---formed either by the active membrane only (\textit{single-layer tube}) or by the coated active membrane (\textit{multilayer tube}). Numerical results showing the influence of the mechanical and the geometrical properties of the soft coating layers on the electromechanical response of the active membrane are presented for different constraint conditions.
\end{abstract}


\maketitle

\section{Introduction}
Dielectric Elastomers (DEs) are a novel class of electro-active
polymers able to change significantly their shape and size when subjected
to an electric stimulus \citep{pelrine1998,pelrine2000,Kofod}. Being lightweight, fast responsive, highly efficient, reliable, and inexpensive, these materials are attractive to be employed in electromechanical transduction technology \citep{Carpi2008,Vertechy2010jimss,Graf2014jimss,RossetShea2016,Ho2017}. Dielectric Elastomer Transducers (DETs) are devised by coating the opposite surfaces of a dielectric elastomer film with stretchable electrodes; as a voltage drop between the electrodes is induced, the dielectric elastomer area expands whereas its thickness shrinks.
As the voltage increases, the elastomeric film thins down inducing in turn a higher electric field; this positive effect may on the contrary result in a catastrophic thinning of the elastomer \citep{ZurloDestrade2017} and lead to the device failure. This phenomenon is know as \textit{electromechanical instability}, and strongly depends on the material model as well as on the boundary conditions \citep{PlanteDubowsky2006,BertoldiGei2011,rudy&gdb11zamp,Zhou2013jimss,puglisi_apl2013,GeiColonnelliSpring2014,rudykh2014prs,Khan2013,AskMenzelRistinmaaPIUTAM,WangChesterJMPS2016,rudykh2017ejmA}.

Among the possible configurations for a dielectric elastomer transducer, the cylindrical one is interesting for different applications \citep{DEtube4braille,DEtube_fiber_actuator}.
\citet{Pipkin66} and \citet{Carpi2004tube} analyzed the electromechanical behaviour of soft dielectric tubes for finite and small strains, respectively.
\citet{Goulbourne2009tube} developed and validated through comparison with experimental data a numerical model to describe axisymmetric deflections of a tubular DE sensor attached to a McKibben actuator.
\citet{Suo2010tube} analyzed the electromechanical instability of a neo-Hookean soft dielectric tube prestretched by a load and actuated by a radial electric field. The critical strain actuation is computed in terms of different design parameters and effects of strain stiffening on the tube response are also considered.
\citet{Zhou2014sms} investigated the electromechanical response of a DE tube actuator with and without boundary constraints. Unconstrained tubes suffer from electromechanical instability which can be avoided by fixing either the axial length (\textit{axial constraint}) or the outer radius (\textit{radial constraint}) of the tube.
\citet{lu2015jmps} explored the critical and post-bulging bifurcation of a dielectric
elastomer tube subjected to electro-mechanical loading, providing a theoretical prediction and an experimental verification of global instability under force or voltage control and localized instability under volume or charge control.
\citet{OgdenMelnikov2016} examined the response of a tubular actuator when subjected to the combination of a radial electric field, an internal pressure and an axial load by employing the fully nonlinear theory of electroelasticity, considering both thick- and thin-walled assumptions.

Recently, several papers dealing with soft electro-active tubes activated via a radial electric field and their applications have been published. \citet{Zhang2017jmps} proposed the use of guided circumferential waves for the ultrasonic non-destructive on-line SHM to detect defects or cracks and for the self-sensing of the actual state of the tube. 
\citet{Cohen2017ijss} investigated the properties of a tube made up of stacked cylindrical dielectric layers separated by flexible electrodes, showing that increasing the number of layers in a stacked cylindrical actuator may lead to instabilities even if the outer radius is fixed.
\citet{BortotShmuel2017sms} proposed a smart system based on array of soft dielectric tubes for tuning sound. \citet{DorfmannOgden2017} investigated the effects on the tube electroelastic response of deformation dependent permittivity. Furthermore, \citet{BortotShmuel2018ijes} showed that dielectric tubes can be prone to diffuse mode instability, while considered stable with respect to global electromechanical instability. 

Nonetheless, tube-like actuators are difficult to realize and, from a practical viewpoint, possible applications for dielectric elastomer transducers suggest the need to ensure proper electrode protection from aggressive agents and electrical safety of the users. To this end, passive layers coating the active membrane can provide opportune insulation. This is an important issue that has began to gain attention in the last years, since the safety of the users and the service life of the electromechanical transducers are decisive factors for the production and commercialization of such devices. Furthermore, often non-perfectly compliant behaviour of the electrodes has to be accounted for, as well as the interaction of the electro-active transducer with a soft actuated body.
\citet{tesi_calabrese} studied a wearable bandage based on a dielectric elastomer actuator and capable of dynamically modulating the pressure exerted on the limbs. The bandage consists of two active layers embedded between two soft passive layers; experimental and numerical investigations showed that the passive layers play a crucial role in transmitting the actuation from the active layers to the load.
\citet{Chen2016} investigated interactions between a dielectric elastomer balloon actuator and an actuated soft body, when the balloon actuator is either embedded inside
pushing the soft body or wrapped outside pulling the soft body.
\citet{Bortot2017JMPS} analyzed, with focus on the stability, the electromechanical response of a multilayer electro-active spherical balloon devised by embedding the active balloon between two protective soft layers. These works show that the geometrical and mechanical properties of the passive layer strongly modify the multilayer system response.

Interfaces play a crucial role in determining the electromechanical response of multilayer electro-active systems. A single-layer electro-active tube is a membrane coated with compliant electrodes, the application of coatings to the active membrane implies the presence of further interfaces. The interface failure shall take place at the electrodes either at the passive layer side, or at the active layer one. The investigation of this kind of failure requires a more in-depth knowledge of the processes occurring at the interfaces of such multilayer systems, and it is left to future works.
Throughout this work, we make the simplifying assumption of perfectly bonded interfaces.

This paper aims to investigate the electromechanical response of soft multilayer electro-active tubes under different constraints, for which the electric actuation results in an expansion of the tube cavity. Focus is placed on understanding how the passive layers influence the nonlinear electro-elastic response of the active tube with their mechanical and geometrical characteristics.

The paper is structured as follow. After having recalled in the first Section the main equations governing the nonlinear electrostatic deformation of heterogeneous soft dielectrics, the modeling of the electromechanical response of electro-active tubes under different constraints is presented. Specifically, axially and radially constrained tubes are considered alongside unconstrained tubes. Two subsections are dedicated to examine the electromechanical behaviour of single-layer and multilayer electro-active tubes, respectively. Results of a numerical investigation aiming to show the influence of the passive layers on the electromechanical response of the multilayer tube are then presented and specialized to a commercially available dielectric elastomer. Concluding remarks are finally provided.

\section{Nonlinear electro-elasticity}\lb{Sec:finite_electro-elasticity}

Following the approach by \citet{Maugin,dorf&ogde05acmc,mcmeeking,suo}, we summarize in this Section the equations governing the nonlinear deformation of heterogeneous soft dielectrics, adopting the standard notation of continuum electromechanics. Throughout this work, quasi-static electromechanic conditions are assumed.

Consider an electro-elastic body consisting of $n$ homogeneous phases perfectly bonded, and occupying the volume $\mathcal{B}^0=\cup_{i=1}^n \mathcal{B}_i^0 \subset \mathbb{R}^3$ in its undeformed configuration. The external boundary $\partial \mathcal{B}^0$ separates the multi-phase body from the surrounding vacuum. Inside the body, a generic interface between two phases $(j)$ and $(k)$ ($j,k=1,...,n$) is denoted as $\partial \mathcal{B}^0_{in (j,k)}$.
When subjected to electrical and/or mechanical loads, the heterogeneous soft dielectric body deforms.
The deformation is described by a sufficiently smooth function $\bchi (\bX)$, mapping a reference point $\bX$ in the undeformed configuration $\mathcal{B}^0$ to its deformed position $\bx=\bchi (\bX)$ in the current configuration $\mathcal{B}$.
Locally, the deformation is described by the deformation gradient tensor with respect to the reference configuration $\mathcal{B}^0$, defined as $\bF = \Grad  \bchi$.
A measure of the deformation is provided by the right and left Cauchy-Green strain tensors, defined as $\bC =\bF^\rmT \bF$ and $\bB =\bF \bF^\rmT$, respectively.
The volume change is given by the quantity $J= \det \bF$. Hence, an incompressible material is constrained to $J=\det \bF=1$.

The current electrostatic state of the dielectric is defined by the electric field $\bE$ and the electric displacements $\bD$.
Under the hypotheses of electrostatics, the local form of Maxwell equations with respect to the current configuration $\mathcal{B}$ takes the form
\beq\lb{MaxwellEq:actual}
\curl\bE=\textbf{0} , \quad \divv\bD=0.
\eeq
The electric field is conservative, according to the first of equations (\ref{MaxwellEq:actual}); therefore, it can be derived as the gradient of an electrostatic potential $\varphi$, namely $\bE(\bx)=-\grad \varphi (\bx)$.

For quasi-static deformations, the balance of linear momentum with respect to the current configuration $\mathcal{B}$ reads
\beq\lb{balance_lin_mom}
\divv \bsigma+\rho \bff={\bf 0},
\eeq
where $\rho$ is the current mass density, $\bff$ is the mechanical force and $\bsigma$ is the symmetric \textit{total stress tensor} incorporating both mechanical and electrical stresses.
In absence of mechanical body force, Eq.~(\ref{balance_lin_mom}) reduces to the equilibrium equation in the current configuration $\mathcal{B}$
\beq\lb{eq:equilibrium}
\divv \bsigma={\bf 0}\, .
\eeq

On the outer boundary of the body $\partial\mathcal{B}$, the electromechanical fields have to satisfy the following jump conditions 
\beq\lb{Jump:outer_surf}
	(\bsigma-\bsigma^\star) \bn =\bt_m, \quad (\bD-\bD^\star) \cdot \bn = -\omega_e, \quad (\bE-\bE^\star) \times \bn = \b0.	
\eeq
Here, $\bn$ is the outward current unit normal vector, $\bt_m$ is a prescribed mechanical traction, $\omega_e$ is the surface charge density, $\bD^\star$ is the outer electric displacement field, and $\bsigma^\star$ is the Maxwell stress outside the body defined in terms of the outer electric field $\bE^\star$ as
\beq\lb{Maxwell_stress}
\bsigma^\star=\epsilon_0 \left[ \bE^\star \otimes \bE^\star- \frac{1}{2}(\bE^\star \cdot \bE^\star)\bI \right],
\eeq
where $\bI$ is the identity tensor and $\epsilon_0$ is the vacuum permittivity (8.854 pF/m).

Across an internal charge-free boundary $\partial\mathcal{B}_{in (j,k)}$ between phases (\textit{j}) and (\textit{k}), the electromechanical fields must fulfill the following jump conditions
\beq\lb{Jump:in_surf}
\salto{0.1} {\bsigma} \bn=  \b0, \quad \salto{0.1} {\bD} \cdot \bn = 0, \quad \salto{0.1} {\bE} \times \bn = \b0.
\eeq
Here, $\salto{0.1} {\bullet} = (\bullet)^{(j)} - (\bullet)^{(k)}$ denotes the jump operator, and  the outward current unit normal vector $\bn$ points from phase (\textit{j}) towards phase (\textit{k}).

In general, since the deformed configuration is a priori unknown, it is convenient to reformulate the electro-elastic problem in Lagrangian description by employing appropriate \textit{pull-back} operations. In this way, we denote the total first Piola-Kirchhoff stress, the nominal electric displacement and electric fields, respectively, as 
\beq\lb{pull-back}
\bP= J \, \bsigma \bF^{-\rmT}, \quad \bE^0 = \bF^{\rmT} \bE, \quad \bD^0= J \bF^{-1} \bD.
\eeq
Therefore, the governing electromechanical equations (\ref{MaxwellEq:actual}) and (\ref{eq:equilibrium}) turn into
\beq\lb{lagrangian:gov_eqs}
\Curl\bE^0=\textbf{0} , \quad \Div\bD^0=0 , \quad \Div \bP={\bf 0}.
\eeq
Accordingly, the jump conditions on the outer boundary of the body (\ref{Jump:outer_surf}) become 
\beq\lb{lagrangian:Jump:outer_surf}
	(\bP-\bP^\star) \bn^0=\bt_M, \quad (\bD-\bD^{0\star}) \cdot \bn^0 = -\omega_E, \quad (\bE^0-\bE^{0\star}) \times \bn^0 = \b0,
\eeq
where $\bt_M \d A=\bt_m \d a $, $\omega_E \d A=\omega_e \d a$ and $\bn^0$ is the outward referential unit normal vector.
The jump conditions across an internal charge-free surface (\ref{Jump:in_surf}) read 
\beq\lb{lagrangian:Jump:in_surf}
\salto{0.1} {\bP} \bn^0=  \b0, \quad \salto{0.1} {\bD^0} \cdot \bn^0 = 0, \quad \salto{0.1} {\bE^0} \times \bn^0 = \b0.
\eeq

For an incompressible material, following \citet{dorf&ogde05acmc}, we derive the total first Piola-Kirchhoff stress and the nominal electric field from an \textit{augmented} energy density function $W(\bF,\bD^0)$ as
\beq\lb{constitutive:eqs}
\bP=\frac{\partial W}{\partial \bF}-p_0 \bF^{-\rmT}, \qquad \bE^0=\frac{\partial W}{\partial \bD^0}.
\eeq
Here, the incompressibility constraint is introduced via the arbitrary \textit{Lagrange multiplier} $p_0$.

\section{Finite deformations of electro-active tubes under different constraints}\lb{Sec:DEtube}

In this section, the electromechanical response of soft dielectric tubes is investigated with respect to different constraints. Specifically, we consider (\textit{i}) axially constrained, (\textit{ii}) radially constrained, and (\textit{iii}) unconstrained tubes. To begin with, following the approach proposed by \citet{OgdenMelnikov2016}, we model the behaviour of electro-active tubes made up of single active membrane, that is \textit{single-layer tubes}. Then we specialize the model to \textit{multilayer tubes}, obtained by embedding the active membrane between two soft passive layers.

\subsection{Single-layer electro-active tubes}\lb{SubSec:DEtube-single}

\begin{figure*}[!t]
	\begin{center}
		\includegraphics[width= 15 cm]{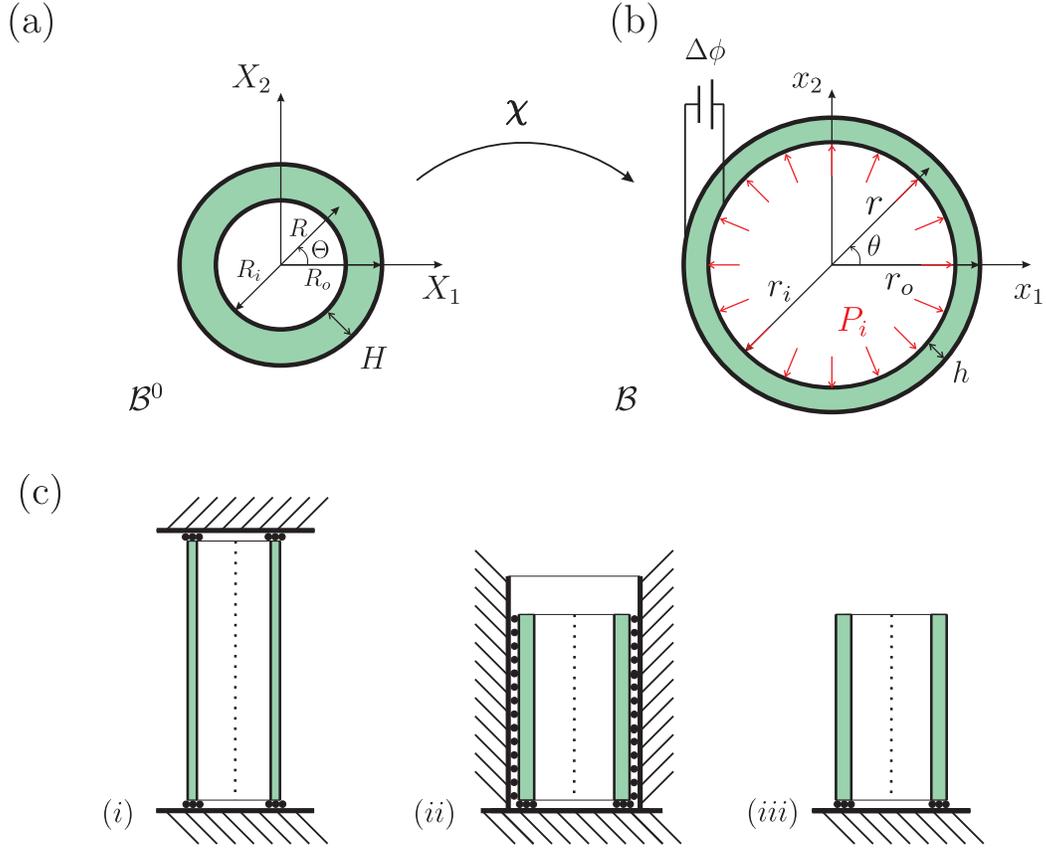}
		\caption{\footnotesize {In-plane cross section of the single-layer electro-active tube in the (a) reference and (b) current configurations. (c) Axial cross sections of the single-layer tube under the different constraints considered, namely (\textit{i}) axially constrained, (\textit{ii}) radially constrained, and (\textit{iii}) unconstrained tubes.}}
		\label{Tube_single}
	\end{center}
\end{figure*}

Consider a thick-walled tube of length much larger than its mean diameter, and made up of an isotropic, hyperelastic and incompressible elastomeric membrane. The inner and outer surfaces of the tube are coated with stretchable electrodes.
In its undeformed configuration $\mathcal{B}^0$, the tube is characterized by inner and outer radii $R_i$ and $R_o$, respectively, so that its initial thickness is $H=R_o-R_i$ (Fig.~\ref{Tube_single}a). Here and hereafter, quantities related to the inner and outer surfaces of the tube are indicated by the notations $(\bullet)_i$ and $(\bullet)_o$, respectively. The tube deforms nonlinearly due to a combination of a pressure $P_i$ applied to its inner surface, and a radial electric field induced by applying an electric potential difference $\Delta \phi$ between the compliant electrodes on its curved surfaces (see Fig.~\ref{Tube_single}b).

Indicating the referential and the current cylindrical coordinate systems, respectively, as $(R,\varTheta, Z)$ and $(r,\theta, z)$, the following mapping describes the deformation undergone by the tube 
\beq\lb{mapping}
r=\sqrt{\lambda_z^{-1}(R^2-R_{i}^2)+r_{i}^2}, \quad \theta=\varTheta, \quad z = \lambda_z Z.
\eeq 
The deformation gradient admits, hence, a diagonal matrix representation
\beq\lb{def_gradient}
\mathsf{F}=\diag\left[(\lambda \lambda_z)^{-1},\lambda,\lambda_z\right],
\eeq
with 
\beq\lb{def:lambda}
\lambda=\frac{r}{R}=\sqrt{\frac{1}{\lambda_z}\left(1-\frac{R_i^2}{R^2}\right)+\lambda_i^2\frac{R_i^2}{R^2}}.
\eeq
Here, $\lambda_{i}=r_{i}/R_{i}$ is the circumferential stretch at the inner surface of the deformed tube. This stretch is mutually related to the stretch at the outer surface, $\lambda_{o}=r_{o}/R_{o}$, as 
\beq\lb{def:lambda-in&out}
\lambda_z \lambda_o^2-1=t^2(\lambda_z \lambda_i^2-1),
\eeq
where $t=R_{i}/R_{o}$ is the \textit{radius ratio} of the tube, such that $0<t<1$ \footnote{The limit cases $t\rightarrow0$ and $t\rightarrow1$ correspond to a cylindrical cavity in an infinite medium and to a thin-walled shell, respectively}.

Due to the finite extensibility of the polymeric chains, dielectric elastomer tubes exhibit a strain-stiffening behaviour and their constitutive response is, thus, well described by a Gent strain-energy function \citep{Gent,Horgan2007,Horgan2015}. For an ideal dielectric elastomer, the electro-elastic strain-energy function reads
\beq\label{strain_energy}
W(\bF,\bD^0)=-\frac{\mu J_m}{2}\left[1-\frac{\tr(\bF^\rmT\bF)-3}{J_m}\right]+\frac{1}{2\epsilon} \bF \bD^0\cdot\bF \bD^0.
\eeq
Here, $\mu$ is the shear modulus of the material, $\epsilon=\epsilon_0\epsilon_r$ is the strain independent permittivity being $\epsilon_r$ the relative dielectric constant, and $J_m$ is the dimensionless locking parameter. In the limit $J_m \rightarrow \infty$ the neo-Hookean model is recovered.

The total stress resulting from Eq.~(\ref{strain_energy}) is
\beq\lb{total_stress}
\bsigma=\frac{\mu}{1-\frac{\tr(\bF^\rmT\bF)-3}{J_{m}}}\bB+\frac{1}{\epsilon}\bD\otimes\bD-p_0\bI,
\eeq
and in the neo-Hookean limit it turns to $\bsigma=\mu \bB+1/\epsilon\bD\otimes\bD-p_0\bI$.

The tube is electrically actuated by connecting the electrodes coating its curved surfaces to a battery. In this way, the voltage (electric potential difference) $\Delta \phi= \varphi(r_o)-\varphi(r_i)$ is applied through the tube thickness, and equal opposing charges accumulate on the electrodes. As a consequence, a radial electric field is induced between the electrodes, $E_r =-\partial \varphi / \partial r$. In terms of the charge per referential unit length, $q=Q/L$, the current electric displacement and electric field read
\beq\lb{electric:disp}
D_r=\frac{q}{2 \pi r \lambda_z}, \qquad E_r=\frac{D_r}{\epsilon}=\frac{q}{2 \pi \epsilon r \lambda_z}.
\eeq
By integration of the second of Eqs.~(\ref{electric:disp}), a relationship connecting the applied voltage and the charge
\beq\lb{voltage:charge}
\Delta \phi=\frac{q}{2 \pi \epsilon \lambda_z} \ln\frac{r_o}{r_i}.
\eeq
Thus, in terms of the applied voltage, the radial electric field reads 
\beq\lb{electric:field}
E_r=\frac{\Delta \phi}{r \ln\frac{r_o}{r_i}}.
\eeq
For the geometry here considered, fringe effects can be neglected and thus, by Gauss's theorem, outside the tube no electric field arise.

For the deformation gradient matrix (\ref{def_gradient}), the strain-energy function (\ref{strain_energy}) depends on the two stretches $\lambda$ and $\lambda_z$ and the following relationships between the stresses can be inferred
\beq\lb{cyl:equ:stress}
\sigma_{\theta \theta}-\sigma_{rr}=\lambda \frac{\partial W(\lambda,\lambda_z)}{\partial \lambda}, \quad \sigma_{zz}-\sigma_{rr}=\lambda_z \frac{\partial W(\lambda,\lambda_z)}{\partial \lambda_z}.
\eeq

Due to the cylindrical symmetry of the deformation, the equilibrium equation (\ref{eq:equilibrium}) reduces to
\beq\lb{equilibrium_eq}
\frac{\d \sigma_{rr}}{\d r}+\frac{\sigma_{rr}-\sigma_{\theta \theta}}{r}=0.
\eeq
Integration of Eq.~(\ref{equilibrium_eq}) provides
\beq\lb{sol:equ:r}
	\sigma_{rr}(r_o)-\sigma_{rr}(r_i)=\int_{r_i}^{r_o}\frac{\sigma_{\theta\theta}-\sigma_{rr}}{r}\d r = \int_{r_i}^{r_o} \lambda \frac{\partial W(\lambda,\lambda_z)}{\partial \lambda}  \frac{\d r}{r}.
\eeq
The integral on the right-hand side of Eq.~(\ref{sol:equ:r}) can be solved using the following change of variable
\beq\lb{change_var}
\frac{\d r}{r} =-\frac{1}{\lambda_z \lambda^2-1}\frac{\d \lambda}{\lambda}.
\eeq

The tube is assumed to have opened ends, and the axial force acting at the ends of the tube is given by
\beq\lb{axial:force}
F_z=\int_{r_i}^{r_o}2 \pi \sigma_{zz}r\d r.
\eeq
%
%
The equilibrium equation (\ref{equilibrium_eq}) along with the first of Eqs.~(\ref{cyl:equ:stress}) gives the following expression for the radial component of the stress
\beq\lb{sigma:rr}
\sigma_{rr} = \frac{1}{2}\left[ \frac{1}{r} \frac{\d}{\d r}(r^2 \sigma_{rr})-\lambda \frac{\partial W(\lambda,\lambda_z)}{\partial \lambda} \right].
\eeq
Obtaining $\sigma_{zz}$ from the second of Eqs.~(\ref{cyl:equ:stress}) and substituting it along with Eq.~(\ref{sigma:rr}) into Eq.~(\ref{axial:force}), we finally obtain the following expression for the axial force
\beq\lb{Fz}
	F_{z}= \pi r^2 \sigma_{rr}|_{r_i}^{r_o} +\pi \int_{r_i}^{r_o} \left(2\lambda_z \frac{\partial W(\lambda,\lambda_z)}{\partial \lambda_z}-\lambda \frac{\partial W(\lambda,\lambda_z)}{\partial \lambda} \right) r \d r.
\eeq
The integral on the right-hand side of Eq.~(\ref{Fz}) can be solved using the following change of variable
\beq\lb{change_var2}
r \d r = - R_i^2\frac{\lambda(\lambda_z \lambda_i^2-1)}{(\lambda_z \lambda^2-1)^2} \d \lambda.
\eeq

Once the boundary conditions are given, Eqs.~(\ref{sol:equ:r}) and (\ref{Fz}) provide the electromechanical response of the electro-active tube.
Next we consider different constraints for the tube. In all the examined cases, a pressure $P_i$ is applied at the inner surface of the tube, so that $\sigma_{rr}(r_i)=-P_i$.

When the tube is \textit{axially constrained} (Fig.~\ref{Tube_single}c(\textit{i})), the tube is clamped at a fixed axial stretch ratio $\lambda_z=\tilde\lambda_z$. The outer surface of the tube is stress free, $\sigma_{rr}(r_o)=0$. From Eq.~(\ref{def:lambda-in&out}), it is thus possible to establish a relationship between $\lambda_o$ and $\lambda_i$
\beq\lb{axial_constraint:lambda-in&out}
\lambda_o^2=\frac{t^2}{\tilde\lambda_z}(\tilde\lambda_z \lambda_i^2-1) + \frac{1}{\tilde\lambda_z}.
\eeq

Integration of Eq.~(\ref{sol:equ:r}), accounting for the boundary conditions, yields
\beq\lb{P:axial_constraint}
\begin{split}
	P_i=&\frac{\mu J_m}{4}\Big[ \ln\frac{\lambda_o}{ \lambda_i}+ \left(1-\frac{\kappa-2}{\sqrt{(\kappa+2)(\kappa-2)}}\right)\times \ln\frac{2\lambda_i^2\tilde\lambda_z-\kappa-\sqrt{(\kappa+2)(\kappa-2)}}{ 2\lambda_o^2\tilde\lambda_z-\kappa-\sqrt{(\kappa+2)(\kappa-2)}} \\
	&+\left(1+\frac{\kappa-2}{\sqrt{(\kappa+2)(\kappa-2)}}\right)\times \ln\frac{2\lambda_i^2\tilde\lambda_z-\kappa+\sqrt{(\kappa+2)(\kappa-2)}}{ 2\lambda_o^2\tilde\lambda_z-\kappa+\sqrt{(\kappa+2)(\kappa-2)}}\Big] \\
	& + \epsilon {E_r^0}^2\frac{(t-1)^2(t^2\lambda_i^2-\lambda_o^2)}{2 t^2\lambda_i^2\lambda_o^2 \left[\ln(\frac{\lambda_o}{t \lambda_i})\right]^2},
\end{split}
\eeq
where $\kappa=(3+J_m)\tilde\lambda_z-\tilde\lambda_z^3$ and $E_r^0=\Delta \phi/H$ is the nominal radial electric field.
In the neo-Hookean limit, Eq.~(\ref{sol:equ:r}) turns into
\beq\lb{P:axial_constraint_nh}
	P_i=\frac{\mu}{2}\left(\frac{\lambda_i^2-\lambda_o^2}{\lambda_i^2\lambda_o^2\tilde\lambda_z^2}+\frac{2}{\tilde\lambda_z} \ln \frac{\lambda_i}{\lambda_o}\right) + \epsilon {E_r^0}^2\frac{(t-1)^2(t^2\lambda_i^2-\lambda_o^2)}{2 t^2\lambda_i^2\lambda_o^2 \left[\ln(\frac{\lambda_o}{t \lambda_i})\right]^2}.
\eeq

When the tube is \textit{radially constrained} (Fig.~\ref{Tube_single}c(\textit{ii})), the outer surface of the tube is held fixed, so that $r_o=R_o$. Since the outer radius of the tube does not change, the stretch at the outer surface is $\lambda_o=1$. The radial stress at the outer surface is equal to the unknown pressure exerted by the constraint, $\sigma_{rr}(r_o)=P_r$. The tube deforms freely in the axial direction, $F_z=0$.
Since $\lambda_o$ is fixed, by employing Eq.~(\ref{def:lambda-in&out}), it is possible to establish a relationship between $\lambda_z$ and $\lambda_i$ as
\beq\lb{radial_constraint:stretch_relation}
\lambda_z=\frac{t^2-1}{t^2 \lambda_i^2-1}.
\eeq

From Eq.~(\ref{sol:equ:r}), we can obtain the pressure exerted by the constraint on the outer surface of the tube as
\beq\lb{Reaction:radial_constraint}
\begin{split}
	P_r=&\frac{\mu J_m}{4} \frac{1}{\sqrt{(\beta-2)(\beta+2)}} \Bigg[-4\sqrt{(\beta-2)(\beta+2)}\ln\lambda_i\\ &+(2+\beta-\sqrt{(\beta-2)(\beta+2)})\times \ln\left[\frac{2\gamma-\beta-\sqrt{(\beta-2)(\beta+2)}}{2\gamma\lambda_i^2-\beta-\sqrt{(\beta-2)(\beta+2)}}\right]\\
	&-(2+\beta+\sqrt{(\beta-2)(\beta+2)})\times \ln\left[\frac{2\gamma-\beta+\sqrt{(\beta-2)(\beta+2)}}{2\gamma\lambda_i^2-\beta+\sqrt{(\beta-2)(\beta+2)}}\right]\Bigg] \\
	& + \epsilon {E_r^0}^2\frac{(t-1)^2(t^2\lambda_i^2-1)}{2t^2\lambda_i^2 \left[\ln(\frac{1}{t \lambda_i})\right]^2} - P_i,
\end{split}
\eeq
where $\beta=(3+J_m)\gamma-\gamma^3$ and $\gamma=(1-t^2)/(1-t^2\lambda_i^2)$.
In the neo-Hookean limit, the pressure exerted by the constraint reads
\beq\lb{Reaction:radial_constraint_nh}
\begin{split}
	P_r=&\frac{\mu}{2 \gamma^2} \left[1+\frac{1}{\lambda_i^2(t^2\lambda_i^2-1)}(1-t^2\lambda_i^2+2(t^2-1) \lambda_i \ln \lambda_i) \right] \\
	&+ \epsilon {E_r^0}^2\frac{(t-1)^2(t^2\lambda_i^2-1)}{2t^2\lambda_i^2 \left[\ln(\frac{1}{t \lambda_i})\right]^2} - P_i.
\end{split}
\eeq

Substituting Eq.~(\ref{Reaction:radial_constraint}) into Eq.~(\ref{Fz}) and imposing $F_z$ equal to zero, we obtain the expression for the internal pressure
\beq\lb{P:radial_constraint}
\begin{split}
	P_i=&\frac{t^2}{t^2\lambda_i^2-1} \Bigg\lbrace \frac{\mu J_m(\gamma \lambda_i^2-1)}{2-\beta} \Bigg[-\frac{(t^2-1)(\gamma^3-1)}{t^2 \gamma (\gamma \lambda_i^2-1)}\\
	&-\frac{\beta-2\gamma^3}{\gamma \sqrt{(2+\beta)(2-\beta)}}\Bigg( \arctan\left[\frac{2\gamma-\beta}{\sqrt{(2+\beta)(2-\beta)}}\right] -\arctan\left[\frac{2\gamma\lambda_i^2-\beta}{\sqrt{(2+\beta)(2-\beta)}}\right]\Bigg)\Bigg]\\
	&-\frac{\mu J_m}{4t^2\sqrt{(\beta-2)(\beta+2)}}\Bigg[ -4\sqrt{(\beta-2)(\beta+2)}\ln\lambda_i +(2+\beta-\sqrt{(\beta-2)(\beta+2)})\times\\ &\ln\left[\frac{2\gamma-\beta-\sqrt{(\beta-2)(\beta+2)}}{2\gamma\lambda_i^2-\beta-\sqrt{(\beta-2)(\beta+2)}}\right]-(2+\beta+\sqrt{(\beta-2)(\beta+2)})\times\\
	&\ln\left[\frac{2\gamma-\beta+\sqrt{(\beta-2)(\beta+2)}}{2\gamma\lambda_i^2-\beta+\sqrt{(\beta-2)(\beta+2)}}\right]\Bigg] + \epsilon {E_r^0}^2\frac{(t-1)^2}{t^2}\left[\frac{1}{ \ln(\frac{1}{t \lambda_i})} -\frac{t^2\lambda_i^2-1}{2t^2\lambda_i^2 \left[\ln(\frac{1}{t \lambda_i})\right]^2} \right]  \Bigg\rbrace.
\end{split}
\eeq
In the neo-Hookean limit, the internal pressure reads
\beq\lb{P:radial_constraint_nh}
\begin{split}
	P_i=&\frac{t^2}{t^2\lambda_i^2-1} \Bigg\lbrace \frac{\mu}{2 t^2 \gamma^2} \Big[2\left((t^2-1)(\gamma^3-1)+t^2(\gamma \lambda_i^2-1)\ln\lambda_i\right) -1-\frac{1}{\lambda_i^2(t^2\lambda_i^2-1)}\\
	&\left(1-t^2\lambda_i^2+2(t^2-1)\lambda_i^2\ln\lambda_i\right) \Big]
	+ \epsilon {E_r^0}^2\frac{(t-1)^2}{t^2}\left[\frac{1}{ \ln(\frac{1}{t \lambda_i})} -\frac{t^2\lambda_i^2-1}{2t^2\lambda_i^2 \left[\ln(\frac{1}{t \lambda_i})\right]^2} \right]   \Bigg\rbrace.
\end{split}
\eeq

Finally, when the tube is \textit{unconstrained} (Fig.~\ref{Tube_single}c(\textit{iii})), its outer surface is stress free, $\sigma_{rr}(r_o)=0$, and the tube can freely deform in the axial direction, $F_z=0$. In this case, it is not possible to establish an explicit relationship between $\lambda_z$ and $\lambda_i$. Nevertheless, the electromechanical response of the tube can be obtained numerically, through Eqs.~(\ref{sol:equ:r}) and (\ref{Fz}). 

In the Appendix, the thin-walled approximation for an axially constrained tube is discussed.

\subsection{Multilayer electro-active tubes}\lb{SubSec:DEtube-multi}

\begin{figure}[t]
	\begin{center}
		\includegraphics[width= 14 cm]{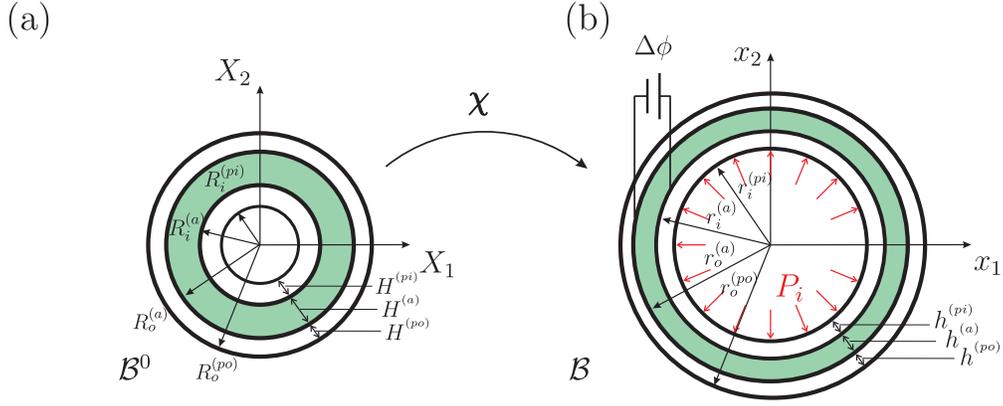}
		\caption{\footnotesize {Cross section of the multilayer electro-active tube in the (a) reference and (b) current configurations.}}
		\label{Tube_multi}
	\end{center}
\end{figure}

Consider next a multilayer tube, devised by coating the active cylindrical membrane with two soft coating passive layers (see Fig.~\ref{Tube_multi}). The surfaces of the membrane to be activated are coated with compliant electrodes; then the active cylindrical membrane is embedded between the two passive layers, resulting thereby insulated from the surrounding space.

Each one of three layers is constituted by an homogeneous, hyperelastic, incompressible material, whose response is described by the strain energy function (\ref{strain_energy}). We assume that the three layers are perfectly bonded. In the reference configuration, the active layer is characterized by inner and outer radii $R_i^{(a)}$ and $R_o^{(a)}$; the inner passive layer is, thus, characterized by outer radii $R_o^{(pi)}=R_i^{(a)}$, and the outer passive layer by inner radii $R_i^{(po)}=R_o^{(a)}$.

Here and in the following, quantities related to the active, inner and outer passive layers are denoted by superscripts $(a)$, $(pi)$ and $(po)$, respectively.

For each layer $(l)$, we define the radius ratio $t^{(l)}=R_i^{(l)}/R_o^{(l)}$. Since at the interface the layers share the same radius, these ratios read
\beq\lb{t_relations_aplayers}
t^{(a)}=\frac{R_i^{(a)}}{R_o^{(a)}}, \quad t^{(pi)}=\frac{R_i^{(pi)}}{R_i^{(a)}}, \quad t^{(po)}=\frac{R_o^{(a)}}{R_o^{(po)}}.
\eeq
In addition, we define the radius ratio of the multilayer tube as $T=R_i^{(pi)}/R_o^{(po)}$. Taking into account the layer radius ratios (\ref{t_relations_aplayers}), the tube radius ratio reads 
\beq
T=\frac{R_i^{(pi)}}{R_o^{(po)}}=t^{(a)} t^{(pi)} t^{(po)},
\eeq 
and it is subjected to the condition $T>0$, so as to exclude the limit case of cylindrical cavity in an infinite medium.
On the basis of Eqs.~(\ref{t_relations_aplayers}), we can define for each layer $(l)$ the reference thickness $H^{(l)}$ as a function of the outer radius of the active layer $R_o^{(a)}$, namely
\beq\lb{H_layers}
H^{(a)}=(1-t^{(a)})R_o^{(a)},\quad H^{(pi)}=(1-t^{(pi)})t^{(a)}R_o^{(a)}, \quad H^{(po)}=\left(\frac{1}{t^{(po)}}-1 \right)R_o^{(a)}.
\eeq
Since the cavity of the cylinder must not vanish, the thickness of the inner passive layer must be smaller than the inner radius of the active membrane, that is $H^{(pi)}<R_i^{(a)}$ and, consequently, $R_i^{(pi)}>0$.

The deformation of the multilayer tube is induced by applying an electric potential difference $\Delta \phi$ across the active layer thickness, and a pressure $P_i$ to the surface of the tube cavity, so that $\sigma_{rr}(r_i^{(pi)})=-P_i$.

Each layer undergoes a deformation according to the mapping (\ref{mapping}).

Since we assume perfect bonding between the layers, at the active-passive layer interfaces, continuity of tractions implies that
\beq\lb{ic:multilayer_balloon}
\sigma_{rr}(r_o^{(pi)})=\sigma_{rr}(r_i^{(a)}),\qquad \sigma_{rr}(r_o^{(a)})=\sigma_{rr}(r_i^{(po)});
\eeq
furthermore, continuity of the displacements requires that
\beq\lb{lambda_relations_aplayers}
\lambda_o^{(pi)}=\lambda_i^{(a)}, \qquad \lambda_i^{(po)}=\lambda_o^{(a)}.
\eeq
Thereby, in light of Eq.~(\ref{def:lambda-in&out}), for each layer the following relationships between the inner and outer stretches can be inferred
\beq\lb{lambda_relations}
\begin{split}
	{\lambda_o^{(a)}}^2-\frac{1}{\lambda_z^{(a)}}&={t^{(a)}}^2\left({\lambda_i^{(a)}}^2-\frac{1}{\lambda_z^{(a)}}\right), \\ {\lambda_i^{(pi)}}^2-\frac{1}{\lambda_z^{(pi)}}&=\frac{1}{{t^{(pi)}}^2}\left({\lambda_i^{(a)}}^2-\frac{1}{\lambda_z^{(pi)}}\right), \\ {\lambda_o^{(po)}}^2-\frac{1}{\lambda_z^{(po)}}&={t^{(po)}}^2\left({\lambda_o^{(a)}}^2-\frac{1}{\lambda_z^{(po)}}\right).
\end{split}
\eeq

Eq.~(\ref{equilibrium_eq}) governs the equilibrium of each layer. Accounting for the interface conditions (\ref{ic:multilayer_balloon}), integrating Eq.~(\ref{equilibrium_eq}) for each layer and summing each contribution, we obtain 
\beq\lb{multilayer:eq:r}
	\sigma_{rr}(r_o^{(po)})-\sigma_{rr}(r_i^{(pi)})=\int_{r_i^{(a)}}^{r_o^{(a)}}\frac{\sigma_{\theta\theta}^{(a)}-\sigma_{rr}^{(a)}}{r}\d r+\int_{r_i^{(pi)}}^{r_o^{(pi)}}\frac{\sigma_{\theta\theta}^{(pi)}-\sigma_{rr}^{(pi)}}{r}\d r +\int_{r_i^{(a)}}^{r_o^{(po)}}\frac{\sigma_{\theta\theta}^{(po)}-\sigma_{rr}^{(po)}}{r}\d r.
\eeq

At th same way the axial force acting at the ends of the multilayer tube is given by the sum of the contribution of each layer
\beq\lb{multilayer:axial:force}
	F_z=\int_{r_i^{(a)}}^{r_o^{(a)}}2 \pi \sigma_{zz}^{(a)}r\d r+\int_{r_i^{(pi)}}^{r_o^{(pi)}}2 \pi \sigma_{zz}^{(pi)}r\d r+\int_{r_i^{(po)}}^{r_o^{(po)}}2 \pi \sigma_{zz}^{(po}r\d r,
\eeq
where $\sigma_{zz}$ in each layer is determined according to the second of Eqs.~(\ref{cyl:equ:stress}) and to Eq.~(\ref{sigma:rr}).

Eqs.~(\ref{multilayer:eq:r}) and (\ref{multilayer:axial:force}) hence define the electromechanical response of the multilayer tube, once the boundary conditions are given. We consider next the different constraints examined for the single-layer tube. For the sake of brevity, we provide the equations governing the electromechanical response of the multilayer tube according to the neo-Hookean model only. 

When the tube is \textit{axially constrained}, the multilayer tube is clamped at a fixed axial stretch ratio $\lambda_z^{(a)}=\lambda_z^{(p)}=\tilde\lambda_z$. The outer surface of the multilayer tube is stress free, $\sigma_{rr}(r_o^{(po)})=0$. Thus, the relationships between the stretches (\ref{lambda_relations}) read
\beq\lb{axial_constraint:lambda-in&out}
\begin{split}
	{\lambda_o^{(a)}}^2-\frac{1}{\tilde\lambda_z}&={t^{(a)}}^2\left({\lambda_i^{(a)}}^2-\frac{1}{\tilde\lambda_z}\right), \\ {\lambda_i^{(pi)}}^2-\frac{1}{\tilde\lambda_z}&=\frac{1}{{t^{(pi)}}^2}\left({\lambda_i^{(a)}}^2-\frac{1}{\tilde\lambda_z}\right), \\ {\lambda_o^{(po)}}^2-\frac{1}{\tilde\lambda_z}&={t^{(po)}}^2\left({\lambda_o^{(a)}}^2-\frac{1}{\tilde\lambda_z}\right).
\end{split}
\eeq
Accounting for the boundary conditions, Eq.~(\ref{multilayer:eq:r}) turns into
\beq\lb{multi:P:axial_constraint_nh}
\begin{split}
	P_i&=\frac{\mu^{(a)}}{2}\left(\frac{{\lambda_i^{(a)}}^2-{\lambda_o^{(a)}}^2}{{\lambda_i^{(a)}}^2{\lambda_o^{(a)}}^2\tlza^2}+\frac{1}{\tlza} \ln \frac{\lambda_i^{(a)}}{\lambda_o^{(a)}}\right) +\frac{\mu^{(pi)}}{2}\left(\frac{{\lambda_i^{(pi)}}^2-{\lambda_i^{(a)}}^2}{{\lambda_i^{(pi)}}^2{\lambda_i^{(a)}}^2\tlzp^2}+\frac{1}{\tlzp} \ln \frac{\lambda_i^{(pi)}}{\lambda_i^{(a)}}\right)\\ &+\frac{\mu^{(po)}}{2}\left(\frac{{\lambda_o^{(a)}}^2-{\lambda_o^{(po)}}^2}{{\lambda_o^{(a)}}^2{\lambda_o^{(po)}}^2\tlzp^2}+\frac{1}{\tlzp} \ln \frac{\lambda_o^{(a)}}{\lambda_o^{(po)}}\right) + \epsilon {E_r^0}^2\frac{(t^{(a)}-1)^2({t^{(a)}}^2{\lambda_i^{(a)}}^2-{\lambda_o^{(a)}}^2)}{2 {t^{(a)}}^2{\lambda_i^{(a)}}^2{\lambda_o^{(a)}}^2 \left[\ln\left(\frac{\lambda_o^{(a)}}{t^{(a)}\lambda_i^{(a)}}\right)\right]^2}.
\end{split}
\eeq

When the tube is \textit{radially constrained}, the outer surface of the multilayer tube is held fixed, so that $r_o^{(po)}=R_o^{(po)}$. Since the outer radius of the outer passive layer does not change, the stretch is fixed $\lambda_o^{(po)}=1$.  The tube can deform freely in the axial direction, $F_z=0$; thus, the axial stretch in the three layers is the same, namely $\lambda_z=\lambda_z^{(a)}=\lambda_z^{(p)}$. Thereby, the relationships between the stretches (\ref{lambda_relations}) become
\beq\lb{axial_constraint:lambda-in&out}
\begin{split}
	{\lambda_o^{(a)}}^2-\frac{1}{\lambda_z}&=\frac{1}{{t^{(po)}}^2}\left(1-\frac{1}{\lambda_z}\right), \\ {\lambda_i^{(a)}}^2-\frac{1}{\lambda_z}&=\frac{1}{{t^{(a)}}^2{t^{(po)}}^2}\left(1-\frac{1}{\lambda_z}\right), \\ {\lambda_i^{(pi)}}^2-\frac{1}{\lambda_z}&=\frac{1}{{t^{(a)}}^2{t^{(po)}}^2{t^{(pi)}}^2}\left(1-\frac{1}{\lambda_z}\right).
\end{split}
\eeq 
The radial stress at the outer surface of the multilayer tube is equal to the unknown pressure exerted by the constraint, $\sigma_{rr}(r_o^{(po)})=P_r$.
From Eq.~(\ref{multilayer:eq:r}), we can obtain the pressure exerted by the constraint on the outer surface of the tube as
\beq\lb{multi:Reaction:radial_constraint_nh}
\begin{split}
	P_r&=\frac{\mu^{(a)}}{2}\left(\frac{{\lambda_i^{(a)}}^2-{\lambda_o^{(a)}}^2}{{\lambda_i^{(a)}}^2{\lambda_o^{(a)}}^2\lz^2}+\frac{1}{\lz} \ln \frac{\lambda_i^{(a)}}{\lambda_o^{(a)}}\right) +\frac{\mu^{(pi)}}{2}\left(\frac{{\lambda_i^{(pi)}}^2-{\lambda_i^{(a)}}^2}{{\lambda_i^{(pi)}}^2{\lambda_i^{(a)}}^2\lz^2}+\frac{1}{\lz} \ln \frac{\lambda_i^{(pi)}}{\lambda_i^{(a)}}\right)\\ &+\frac{\mu^{(po)}}{2}\left(\frac{{\lambda_o^{(a)}}^2-1}{{\lambda_o^{(a)}}^2\lz^2}+\frac{1}{\lz} \ln {\lambda_o^{(a)}}\right) + \epsilon {E_r^0}^2\frac{(t^{(a)}-1)^2({t^{(a)}}^2{\lambda_i^{(a)}}^2-{\lambda_o^{(a)}}^2)}{2 {t^{(a)}}^2{\lambda_i^{(a)}}^2{\lambda_o^{(a)}}^2 \left[\ln\left(\frac{\lambda_o^{(a)}}{t^{(a)}\lambda_i^{(a)}}\right)\right]^2} - P_i.
\end{split}
\eeq
Substituting Eq.~(\ref{multi:Reaction:radial_constraint_nh}) into Eq.~(\ref{multilayer:axial:force}) and equating $F_z$ to zero, we obtain the following expression for the internal pressure
\beq\lb{multi:P:radial_constraint_nh}
\begin{split}
	P_i=&\frac{{t^{(a)}}^2{t^{(po)}}^2}{{t^{(a)}}^2{t^{(pi)}}^2{t^{(po)}}^2-1}\Bigg\lbrace\frac{\mu^{(a)}}{\lz^2}\Bigg[\frac{1}{t^{(a)}}\Big(({t^{(a)}}^2-1)(\lz^3-1)+{t^{(a)}}^2({\lambda_i^{(a)}}^2\lz-1) \ln \frac{\lambda_i^{(a)}}{\lambda_o^{(a)}}\Big)\\
	&-\frac{1}{2}\left(\frac{{\lambda_i^{(a)}}^2-{\lambda_o^{(a)}}^2}{{\lambda_i^{(a)}}^2{\lambda_o^{(a)}}^2}+\lz \ln \frac{\lambda_i^{(a)}}{\lambda_o^{(a)}}\right) \Bigg] +\frac{\mu^{(pi)}}{\lz^2}\Bigg[({t^{(pi)}}^2-1)(\lz^3-1)+{t^{(pi)}}^2({\lambda_i^{(pi)}}^2\lz-1)\times\\
	& \ln \frac{\lambda_i^{(pi)}}{\lambda_i^{(a)}}-\frac{1}{2}\left(\frac{{\lambda_i^{(pi)}}^2-{\lambda_i^{(a)}}^2}{{\lambda_i^{(pi)}}^2{\lambda_i^{(a)}}^2}+\lz \ln \frac{\lambda_i^{(pi)}}{\lambda_i^{(a)}}\right)\Bigg]+ \frac{\mu^{(po)}}{\lz^2}\Bigg[\frac{1}{{t^{(a)}}^2{t^{(po)}}^2}\Bigg(({t^{(po)}}^2-1)(\lz^3-1)
	\\
	&+{t^{(po)}}^2 ({\lambda_i^{(po)}}^2\lz-1) \ln {\lambda_o^{(a)}} \Bigg)- \frac{1}{2} \left(\frac{{\lambda_o^{(a)}}^2-1}{{\lambda_o^{(a)}}^2}+\lz \ln {\lambda_o^{(a)}}\right) \Bigg]
	+ \epsilon {E_r^0}^2 \frac{(t^{(a)}-1)^2}{{t^{(a)}}^2}\times\\
	& \Bigg[ \frac{1}{ \ln\frac{\lambda_o^{(a)}}{t^{(a)}\lambda_i^{(a)}}}-\frac{{t^{(a)}}^2{\lambda_i^{(a)}}^2-{\lambda_o^{(a)}}^2}{2 {\lambda_i^{(a)}}^2{\lambda_o^{(a)}}^2 \left[\ln\left(\frac{\lambda_o^{(a)}}{t^{(a)}\lambda_i^{(a)}}\right)\right]^2}\Bigg]\Bigg\rbrace.
\end{split}
\eeq

Finally, when the tube is \textit{unconstrained}, its outer surface is stress free, $\sigma_{rr}(r_o^{(po)})=0$, and the tube can freely deform in the axial direction, $F_z=0$. In this case, it is not possible to establish an explicit relationship between $\lambda_z$ and $\lambda_i^{(a)}$ or $\lambda_i^{(pi)}$. Nevertheless, the electromechanical response of the tube can be obtained numerically, through Eqs.~(\ref{multilayer:eq:r}) and (\ref{multilayer:axial:force}). 

In the Appendix, the thin-walled approximation for an axially constrained multilayer tube is discussed.

\section{Numerical investigation}\lb{Sec:num_investigations}

The numerical investigation aims to show, for pure electric loading ($P_i=0$)\footnote{Applying an internal pressure is beneficial to reduce the electric load required for the actuation of the tube.}, how the soft passive layers modify the electromechanical response of the active membrane, and how the deformation of the tube cavity changes when the active membrane is coated. 

To this end, for the different constraints considered, we first compare the electromechanical response of a single-layer tube with that of a multilayer one, made up of three identical layers. Then, we focus our attention on how the behavior of the multilayer tube is affected by the shear modulus and the thickness of the passive layers.

For convenience, the following dimensionless parameters are introduced
\beq\lb{norm_par}
\overline{\Delta \phi}=\frac{\Delta \phi}{H}\sqrt{\frac{\epsilon^{(a)}}{\mu^{(a)}}},\;\;\; \widetilde{H}^{(l)}= \frac{H^{(l)}}{R^{(a)}_o}, \;\;\; \alpha^{(pi)}= \frac{\mu^{(pi)}}{\mu^{(a)}}, \;\;\; \alpha^{(po)}= \frac{\mu^{(po)}}{\mu^{(a)}}.
\eeq
We assume the same locking parameter $J_m=50$ for both the active and the passive elastomeric membranes. This value corresponds to a limiting uniaxial stretch equal to 7, which is reasonable for silicone-like dielectric elastomers.

Note that, in all the plots to follow, the results for the Gent and the neo-Hookean models are denoted by continuous and dashed curves, respectively.

\begin{figure*}[!htp]
	\begin{center}
		\includegraphics[width= 11cm]{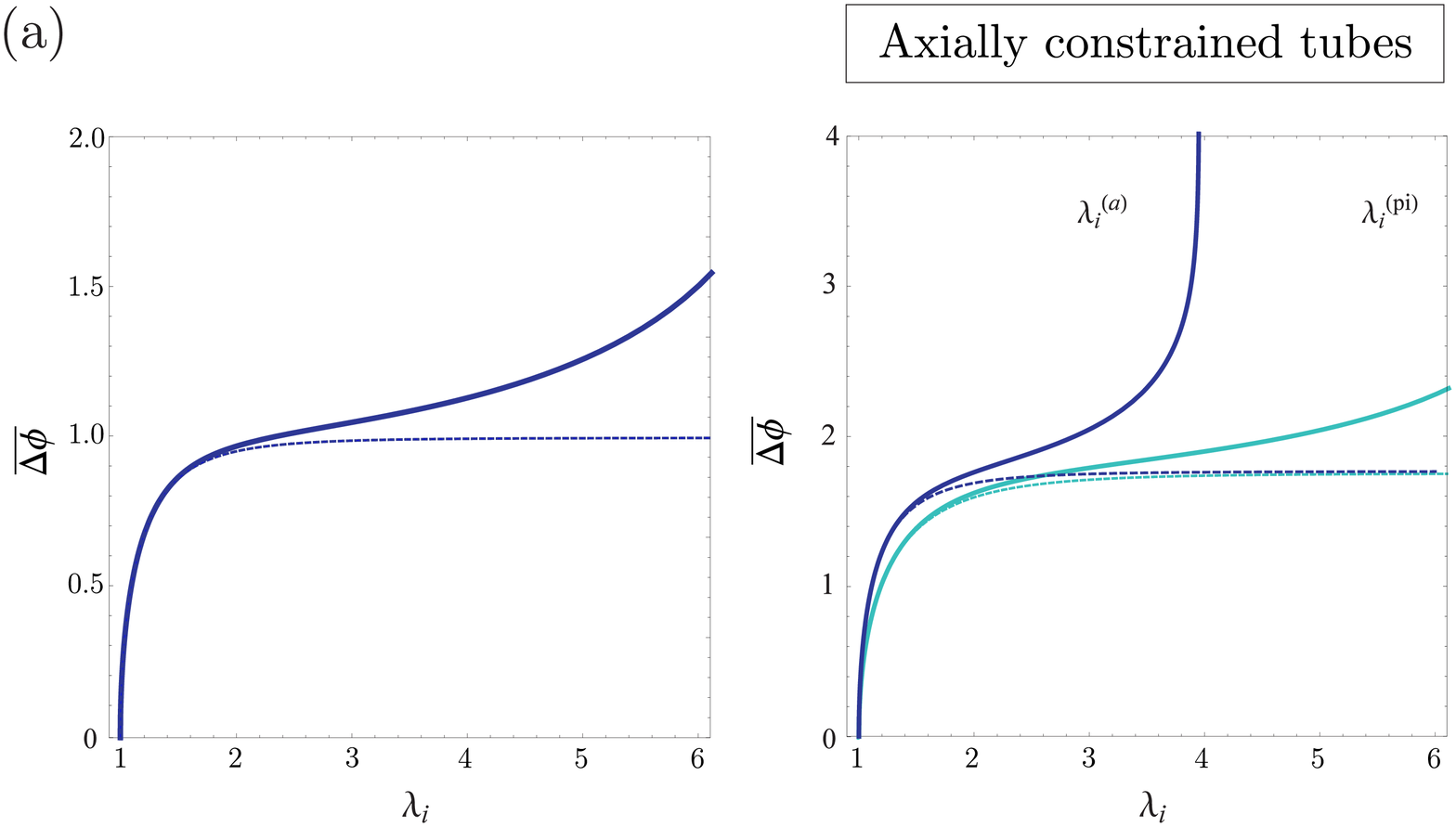}\\
		\vspace{-0.4cm}
		\includegraphics[width= 11cm]{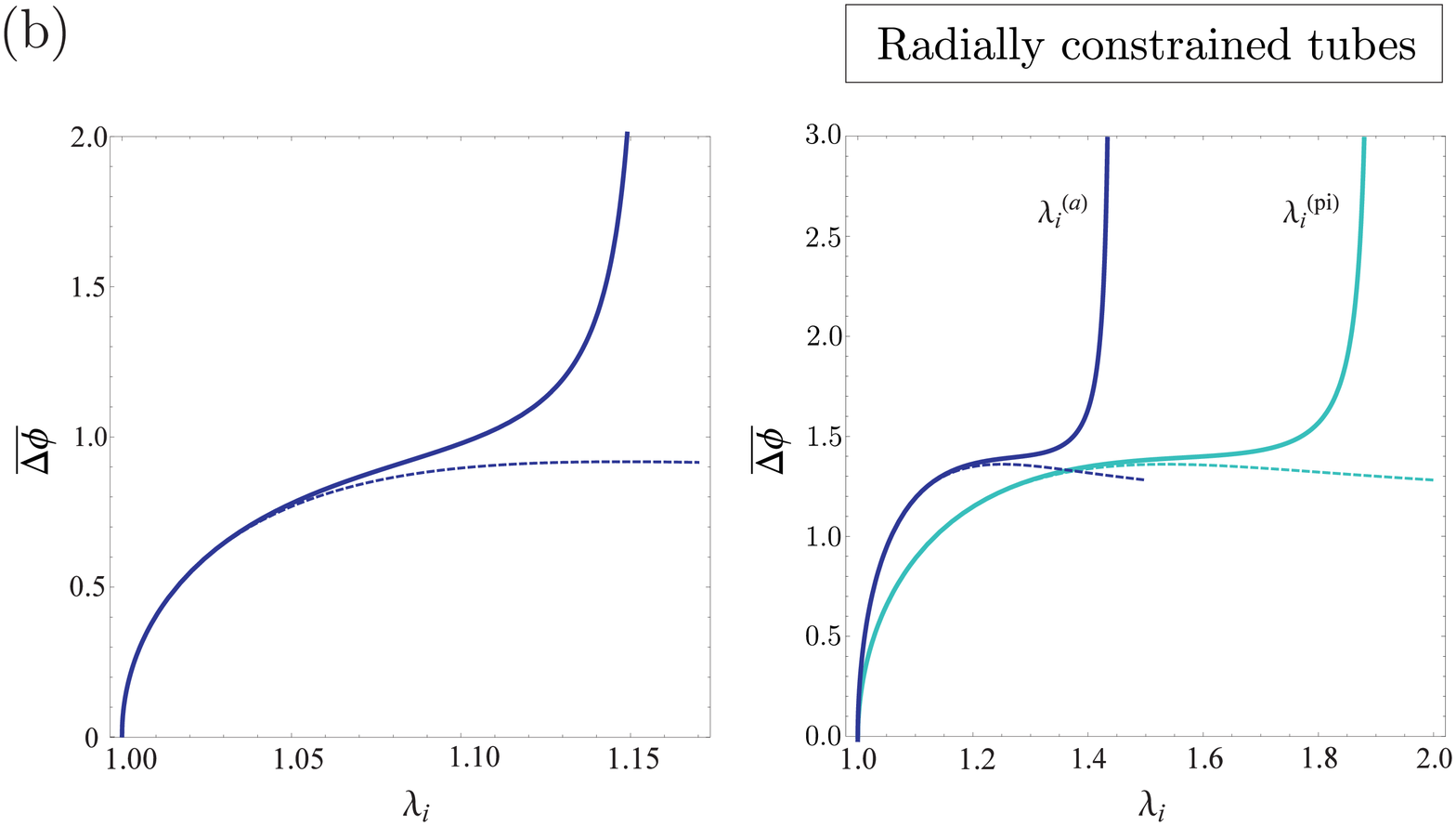}\\
		\vspace{-0.4cm}
		\includegraphics[width= 11cm]{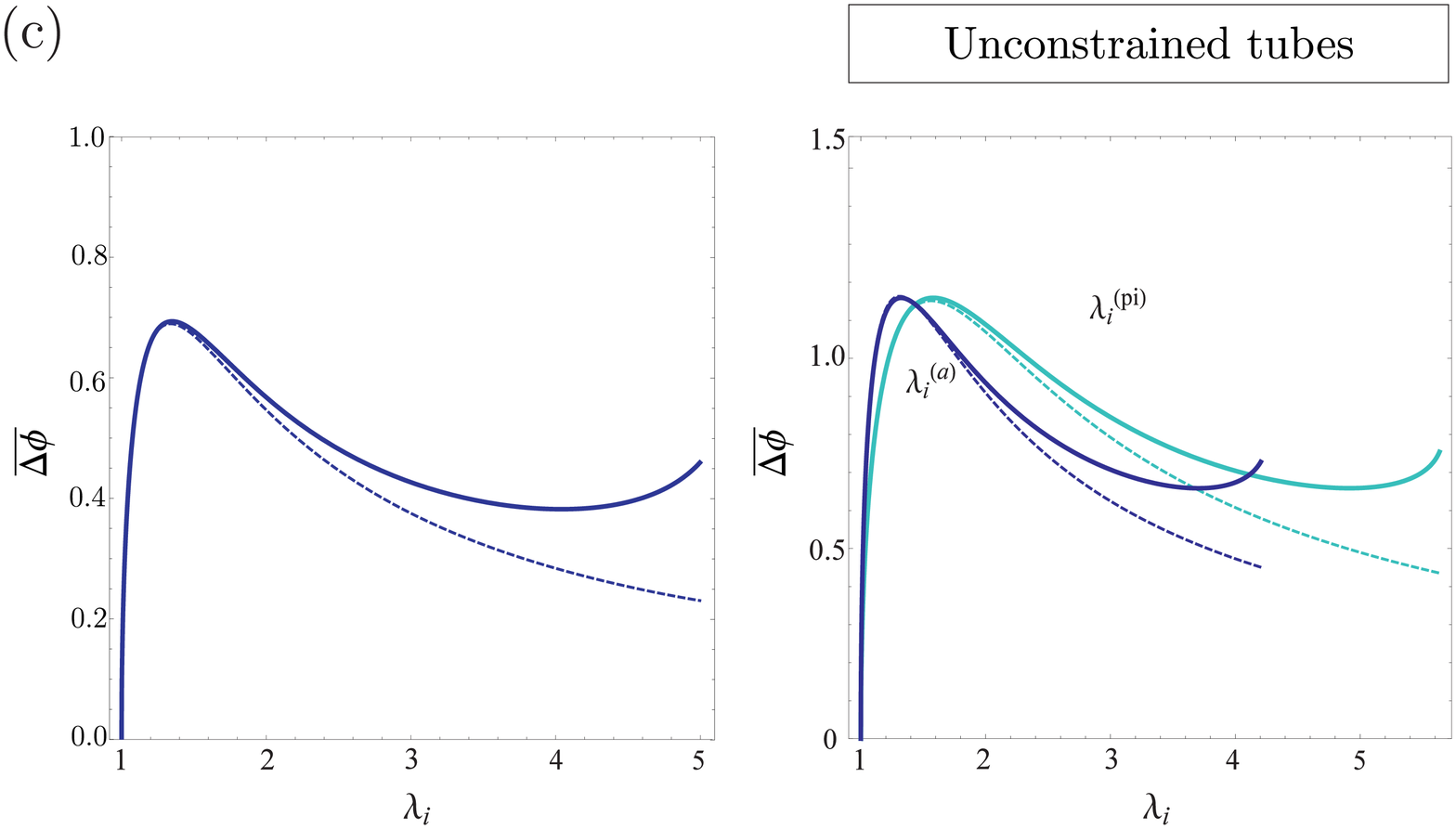}\\
		\caption{\footnotesize {Electromechanical response of soft tubes under pure electric loading ($P_i=0$). Dependence of $\overline{\Delta \phi}$ on the inner circumferential stretch $\lambda_i$ for a single-layer tube (left-hand panels) with active membrane of dimensionless thickness $\widetilde{H}=0.2$, and for a multilayer tube (right-hand panels) obtained by coating the active layer with two identical passive layers, $\widetilde{H}^{(a)}=\widetilde{H}^{(pi)}=\widetilde{H}^{(po)}=0.2$ and $\alpha^{(p)}=\alpha^{(pi)}=\alpha^{(po)}=1$. The tubes are (a) axially constrained at $\tilde{\lambda}_z=1$, (b) radially constrained and (c) unconstrained. Continuous and dashed curves are referred to Gent and neo-Hookean models, respectively.}}
		\label{DEtube_comparison}
	\end{center}
\end{figure*}

\begin{figure*}[!hpt]
	\begin{center}
		\includegraphics[width= 13 cm]{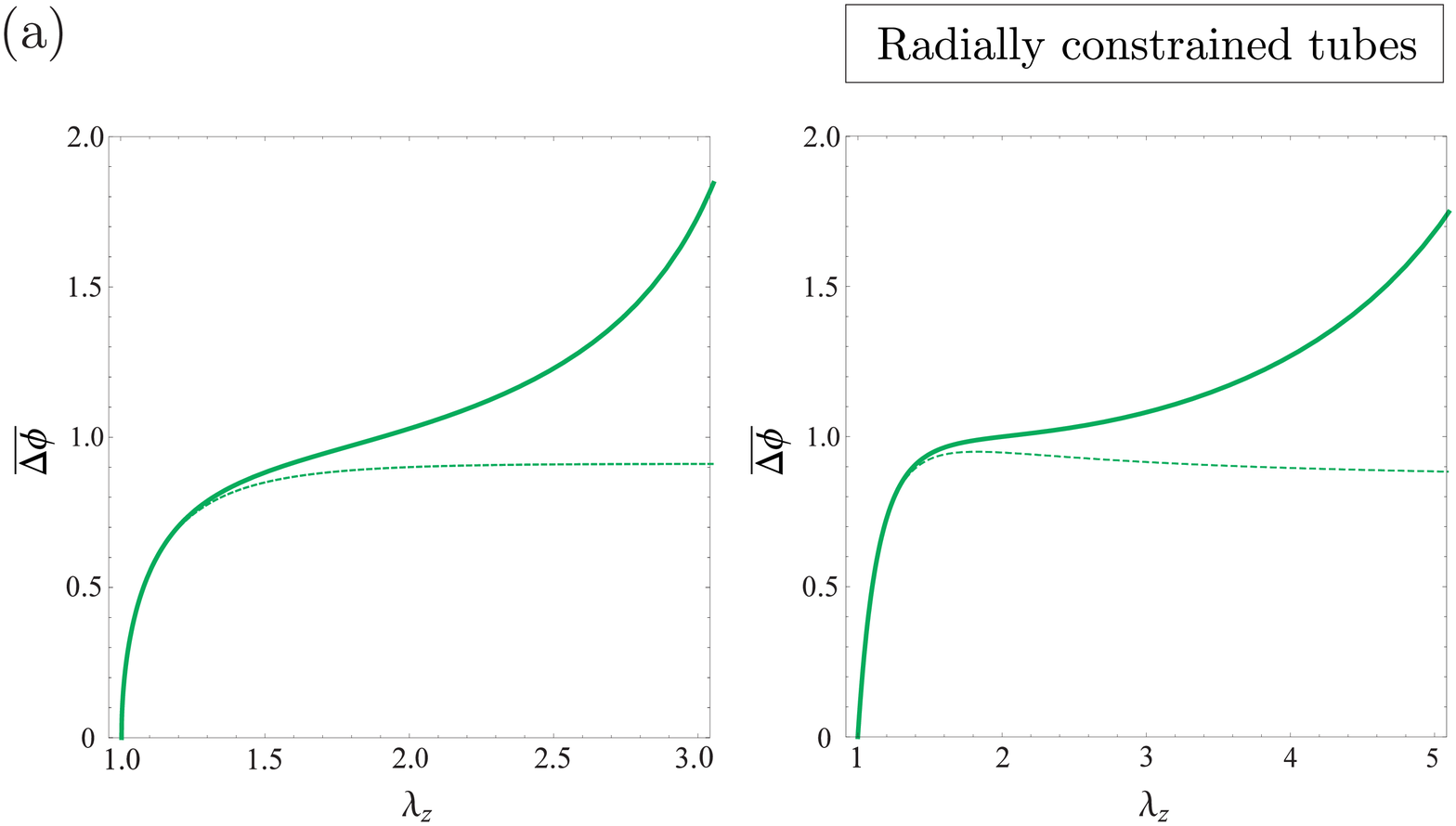}\\
		\vspace{-0.4cm}
		\includegraphics[width= 13 cm]{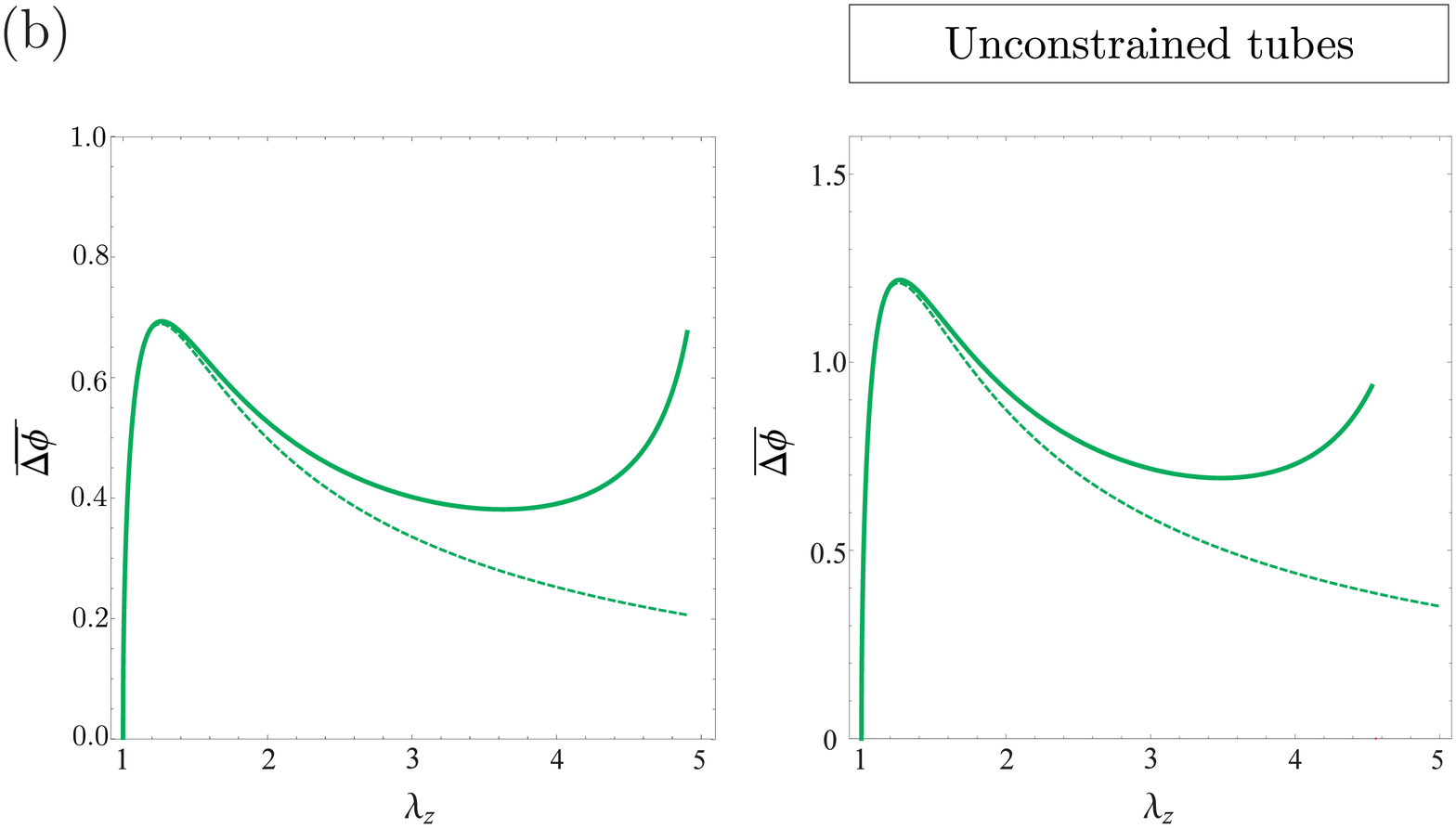}\\
		\caption{\footnotesize {Electromechanical response of soft tubes under pure electric loading ($P_i=0$). Dependence of $\overline{\Delta \phi}$ on the axial stretch $\lambda_z$ for a single-layer tube (left-hand panels) with active membrane of dimensionless thickness $\widetilde{H}=0.2$, and for a multilayer tube (right-hand panels) obtained by coating the active layer with two identical passive layers, such that $\widetilde{H}^{(a)}=\widetilde{H}^{(pi)}=\widetilde{H}^{(po)}=0.2$ and $\alpha^{(p)}=\alpha^{(pi)}=\alpha^{(po)}=1$. The tubes are (a) radially constrained and (b) unconstrained. Continuous and dashed curves are referred to Gent and neo-Hookean models, respectively.}}
		\label{DEtube_comparison-z}
	\end{center}
\end{figure*}

The dependence of $\overline{\Delta \phi}$ on the inner circumferential stretch $\lambda_i$ for a single-layer tube and for a multilayer one, made up of three identical layers, are presented in the left- and right-hand panels of Fig.~\ref{DEtube_comparison}, respectively.

For \textit{axially constrained} tubes (Fig.~\ref{DEtube_comparison}a), the electromechanical response is stable and consists in radial expansion at fixed axial length. In order to obtain the same deformation of the tube cavity, for the multilayer tube a one and a half times higher $\overline{\Delta \phi}$ is required. In the multilayer setting, the deformation of the active membrane is limited due to presence of the passive layers, and its lock-up stretch is smaller.

For \textit{radially constrained} tubes (Fig.~\ref{DEtube_comparison}b), the electromechanical response is stable and consists in axial elongation and radial expansion of the cavity at fixed outer radius. For the multilayer tube, since the constrained outer surface is that of the outer passive layer, the active membrane is more compliant and the deformation of the cavity is larger with respect to the single-layer tube.

For \textit{unconstrained} tubes (Fig.~\ref{DEtube_comparison}c), the electromechanical response is unstable and consists in axial elongation and radial expansion. Increasing the electric load, the tube gradually expands and extends until a critical value of electric load is attained. At this state, a further increase in the electric load may lead to transition of the tube to a new stable state characterized by larger volume of the cavity (\textit{snap-through} instability). Note that this phenomenon, being strictly connected to the strain stiffening, cannot be recovered by the neo-Hookean model. Experimental evidence by \citet{lu2015jmps} confirms that for dielectric elastomer tube electromechanically loaded the response curve is N-shaped, validating hence the Gent model. The same phenomenon occurs for spherical membrane mechanically loaded \citep{Alexander71} as well as for electro-active ballons \citep{keplinger12_balloon_snap,rudy&etal12ijnm}.
With respect to the single-layer tube, for the multilayer one the locking stretch of the active membrane is smaller. The locking stretch at the tube cavity, though, is larger. At the onset of electromechanical instability, the deformation of the active membrane ($\lambda_i^{(a)}=1.32$) is approximately equal to the single-layer case ($\lambda_i=1.35$), although the deformation of the multilayer tube cavity is larger ($\lambda_i^{(pi)}=1.58$). However, the critical $\overline{\Delta \phi}$ for the multilayer tube is three times higher with respect to that for the single-layer one.

For the radially constrained and unconstrained tubes, the dependence of $\overline{\Delta \phi}$  on the axial stretch $\lambda_z$ for a single-layer tube and a multilayer one, made up of three identical layers, are presented in the left- and right-hand panels of Fig.~\ref{DEtube_comparison-z}, respectively.

For \textit{radially constrained} tubes (Fig.~\ref{DEtube_comparison-z}a), to achieve the same axial elongation an approximately double $\overline{\Delta \phi}$ is needed for the multilayer tube. Furthermore, in the multilayer configuration, the maximum stretch attainable at a given voltage according to the Gent model is almost double.

For \textit{unconstrained} tubes (Fig.~\ref{DEtube_comparison-z}b), the axial elongation at the onset of electromechanical instability is the same for both the single-layer and the multilayer tubes, however the critical $\overline{\Delta \phi}$ for the multilayer configuration is one and a half times higher with respect to the single-layer configuration. Furthermore, the lock-up axial stretch of the multilayer tube is smaller.

Comparing the different constraint conditions, we can see that unconstrained tubes requires lower $\overline{\Delta \phi}$ for their actuation, however they are prone to electromechanical instability.
Radially constrained tubes allow for larger axial extension, whereas axially constrained tubes enable a larger cavity expansion. However, an improved deformation of the cavity can be obtained for the multilayer tube.

\begin{figure*}[!ht]
	\begin{center}
		\includegraphics[width= 14 cm]{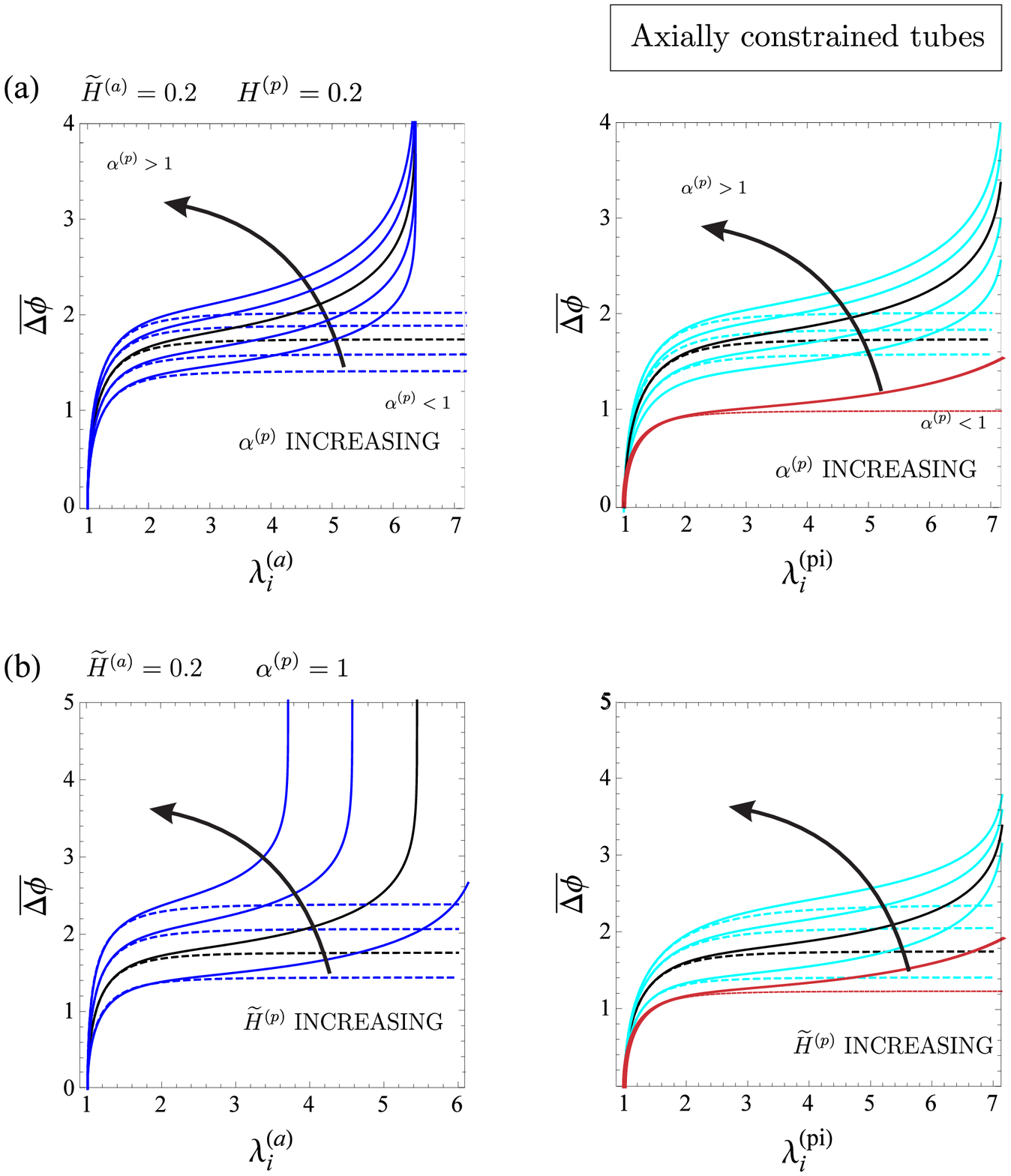}\\
		\caption{\footnotesize {\textit{Axially constrained multilayer tubes}. Electromechanical response of the tube for increasing values of (a) the passive layer shear modulus, namely $\alpha^{(p)}=0.5,0.75,1,1.25,1.5$, and of (b) the passive layer thickness, namely $\widetilde{H}^{(p)}=0.5,1,1.5,2\;\widetilde{H}^{(a)}$. The black curves are associated with the case of identical active and passive layers ($\widetilde{H}^{(a)}=\widetilde{H}^{(pi)}=\widetilde{H}^{(po)}=0.2$ and $\alpha^{(p)}=\alpha^{(pi)}=\alpha^{(po)}=1$). The red curves indicate the deformation of the cavity of the single-layer tube. Continuous and dashed curves are referred to Gent and neo-Hookean models, respectively.}}
		\label{DEtube_Multi:AxConst_DiffPar}
	\end{center}
\end{figure*}

Next, for the different constraint conditions, we examine the effect of modifications in the mechanical or in the geometrical properties of the passive layers on the response of the multilayer tube.

Considering an \textit{axially constrained} multilayer tube formed by an active membrane with $\widetilde{H}^{(a)}=0.2$ and by two identical passive layers ($\widetilde{H}^{(pi)}=\widetilde{H}^{(po)}=\widetilde{H}^{(p)}$ and $\alpha^{(pi)}=\alpha^{(po)}=\alpha^{(p)}$), Fig.~\ref{DEtube_Multi:AxConst_DiffPar} depicts the electromechanical response of the tube for increasing values of (a) the passive layer shear modulus, namely $\alpha^{(p)}=0.5,0.75,1,1.25,1.5$, and of (b) the passive layer thickness, namely $\widetilde{H}^{(p)}=0.5,1,1.5,2\;\widetilde{H}^{(a)}$. 
The black curves are referred to the case of identical active and passive layers ($\widetilde{H}^{(a)}=\widetilde{H}^{(pi)}=\widetilde{H}^{(po)}=0.2$ and $\alpha^{(p)}=\alpha^{(pi)}=\alpha^{(po)}=1$). The red curves indicate the deformation of the cavity of the single-layer tube.

Increasing the shear modulus of the passive layers (Fig.~\ref{DEtube_Multi:AxConst_DiffPar}a) results in a proportional increase of the overall stiffness of the multilayer tube. When the passive layers are softer than the active membrane, the multilayer tube is more deformable and its electromechanical response is hence closer to that of a tube formed by the active membrane only. 

At the same way, increasing the thickness of the passive layers (Fig.~\ref{DEtube_Multi:AxConst_DiffPar}b) results in a proportional increase of the overall stiffness of the multilayer tube. As the passive layers become thicker, the active membrane becomes less deformable and its locking stretch decreases. When the passive layers are thinner than the active membrane, the multilayer tube is more deformable and its electromechanical response is hence closer to that of a tube formed by the active membrane only.

\begin{figure*}[h]
	\begin{center}
		\includegraphics[width= 14 cm]{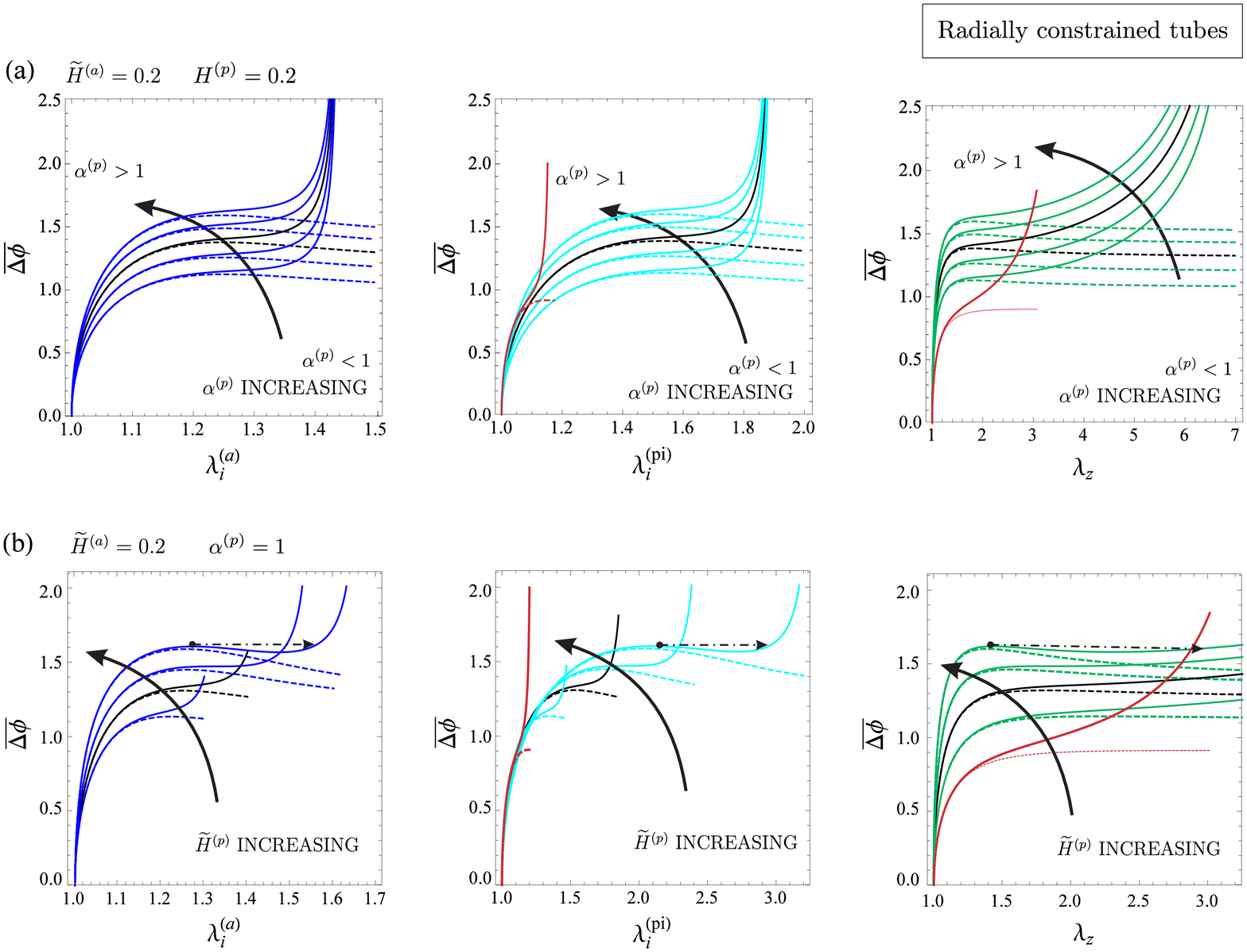}
		\caption{\footnotesize {\textit{Radially constrained multilayer tubes}. Electromechanical response of the tube for different values of (a) the passive layer shear modulus, namely $\alpha^{(p)}=0.5,0.75,1,1.25,1.5$, and of (b) the passive layer thickness, namely $\widetilde{H}^{(p)}=0.5,1,1.5,2\;\widetilde{H}^{(a)}$. The black curves are associated with the case of identical active and passive layers ($\widetilde{H}^{(a)}=\widetilde{H}^{(pi)}=\widetilde{H}^{(po)}=0.2$ and $\alpha^{(p)}=\alpha^{(pi)}=\alpha^{(po)}=1$). The red curves indicate the deformation of the cavity, and the axial deformation of the single-layer tube. Continuous and dashed curves are referred to Gent and neo-Hookean models, respectively.}}
		\label{DEtube_Multi:RadConst_DiffPar}
	\end{center}
\end{figure*}

\begin{figure*}[h]
	\begin{center}
		\includegraphics[width= 14 cm]{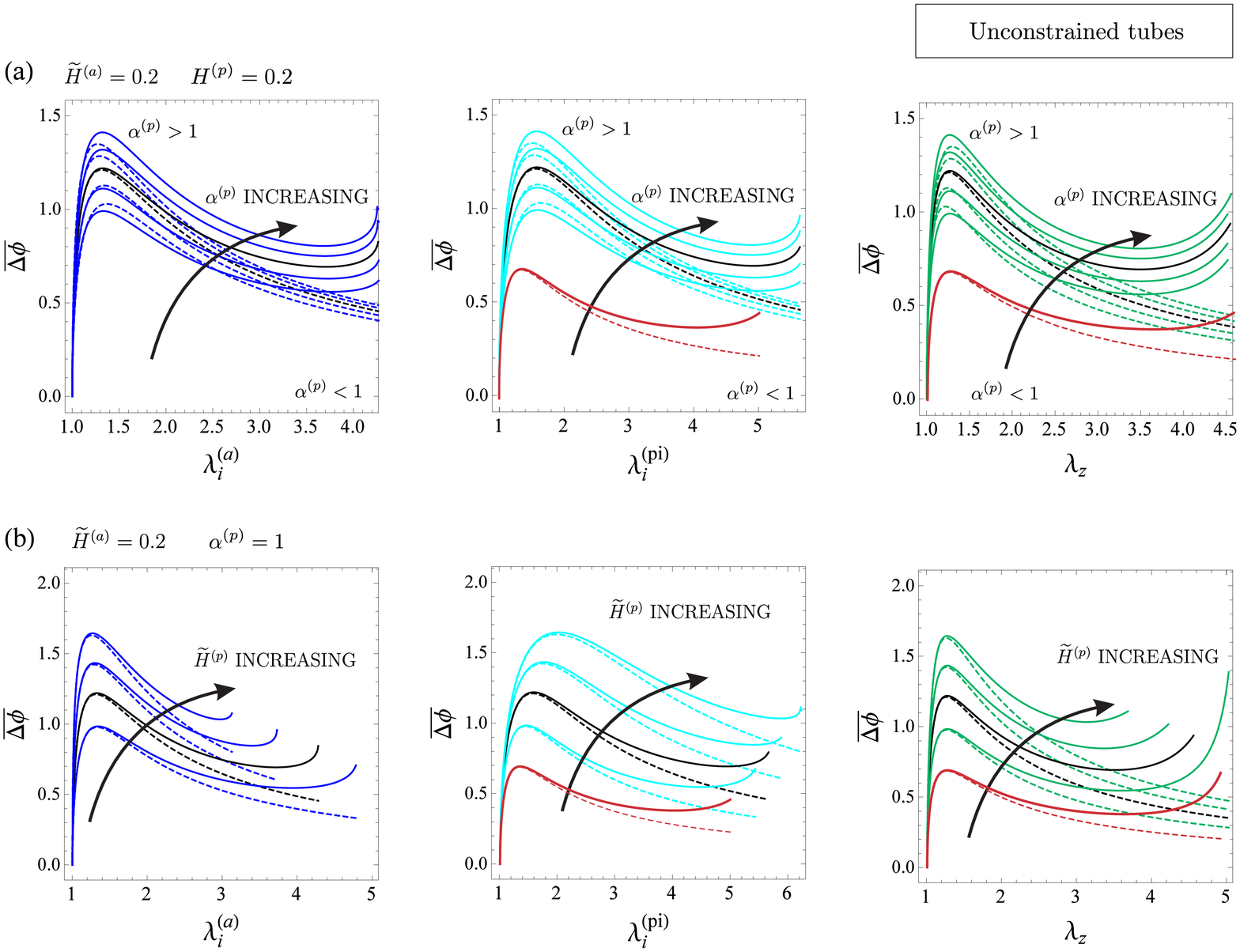}
		\caption{\footnotesize {\textit{Unconstrained multilayer tubes}. Electromechanical response of the tube for increasing values of (a) the passive layer shear modulus, namely $\alpha^{(p)}=0.5,0.75,1,1.25,1.5$, and of (b) the passive layer thickness, namely $\widetilde{H}^{(p)}=0.5,1,1.5,2\;\widetilde{H}^{(a)}$. The black curves are associated with the case of identical active and passive layers ($\widetilde{H}^{(a)}=\widetilde{H}^{(pi)}=\widetilde{H}^{(po)}=0.2$ and $\alpha^{(p)}=\alpha^{(pi)}=\alpha^{(po)}=1$). The red curves indicate the deformation of the cavity, and the axial deformation of the single-layer tube. Continuous and dashed curves are referred to Gent and neo-Hookean models, respectively.}}
		\label{DEtube_Multi:UnConst_DiffPar}
	\end{center}
\end{figure*}

Considering then a \textit{radially constrained} multilayer tube formed by an active membrane with $\widetilde{H}^{(a)}=0.2$ and by two identical passive layers ($\widetilde{H}^{(pi)}=\widetilde{H}^{(po)}=\widetilde{H}^{(p)}$ and $\alpha^{(pi)}=\alpha^{(po)}=\alpha^{(p)}$), Fig.~\ref{DEtube_Multi:RadConst_DiffPar} shows the electromechanical response of the tube for increasing values of (a) the passive layer shear modulus, namely $\alpha^{(p)}=0.5,0.75,1,1.25,1.5$, and of (b) the passive layer thickness, namely $\widetilde{H}^{(p)}=0.5,1,1.5,2\;\widetilde{H}^{(a)}$.
The black curves are referred to the case of identical active and passive layers ($\widetilde{H}^{(a)}=\widetilde{H}^{(pi)}=\widetilde{H}^{(po)}=0.2$ and $\alpha^{(p)}=\alpha^{(pi)}=\alpha^{(po)}=1$). The red curves indicate the deformation of the cavity, and the axial deformation of the single-layer tube.

An increase in the shear modulus of the passive layers (Fig.~\ref{DEtube_Multi:RadConst_DiffPar}a) results in a proportional increase of the electric load required to induce a given deformation. As the passive layers become stiffer, also the overall response of the tube becomes stiffer. Anyway, due to the presence of the outer passive layer, the multilayer tube is more deformable with respect to the single-layer tube. Even strongly decreasing the shear modulus of the passive layers, the electromechanical response of the multilayer tube largely differs from that of the single-layer one.

Increasing the thickness of the passive layers (Fig.~\ref{DEtube_Multi:RadConst_DiffPar}b), the response of the multilayer tube becomes unstable. When the passive layer thickness is twice larger than that of the active membrane ($\widetilde{H}^{(p)}=2\widetilde{H}^{(a)}$) electromechanical instability takes place. As indicated by the dot-dashed arrow in Fig.~\ref{DEtube_Multi:RadConst_DiffPar}b, the tube may experience a snap-through at $\overline{\Delta \phi}=1.69$ from $\lambda_z=1.43$ to $\lambda_z=3.15$. As the thickness of the passive layer decreases, the response of the multilayer tube get closer to that of the single-layer one.

Finally, considering an \textit{unconstrained} multilayer tube formed by an active membrane with $\widetilde{H}^{(a)}=0.2$ and by two identical passive layers ($\widetilde{H}^{(pi)}=\widetilde{H}^{(po)}=\widetilde{H}^{(p)}$ and $\alpha^{(pi)}=\alpha^{(po)}=\alpha^{(p)}$), Fig.~\ref{DEtube_Multi:UnConst_DiffPar} depicts the electromechanical response of the tube for increasing values of (a) the passive layer shear modulus, namely $\alpha^{(p)}=0.5,0.75,1,1.25,1.5$, and of (b) the passive layer thickness, namely $\widetilde{H}^{(p)}=0.5,1,1.5,2\;\widetilde{H}^{(a)}$. 
The black curves are referred to the case of identical active and passive layers ($\widetilde{H}^{(a)}=\widetilde{H}^{(pi)}=\widetilde{H}^{(po)}=0.2$ and $\alpha^{(p)}=\alpha^{(pi)}=\alpha^{(po)}=1$). The red curves indicate the deformation of the cavity, and the axial deformation of the single-layer tube.

Increasing the shear modulus of the passive layers (Fig.~\ref{DEtube_Multi:UnConst_DiffPar}a), the critical value of$\overline{\Delta \phi}$ increases proportionally, while the critical stretch decreases. Contrary to the cases of equal and stiffer passive layers $\alpha^{(p)}\geq1$, when the passive layers are softer than the active membrane $\alpha^{(p)}<1$, the critical value $\overline{\Delta \phi}$ predicted by the neo-Hookean model is higher than that expected according to the Gent model. 

Increasing the thickness of the passive layers (Fig.~\ref{DEtube_Multi:UnConst_DiffPar}b), the critical value of $\overline{\Delta \phi}$ increases proportionally, as well as the critical and the strain-stiffening stretches at the cavity surface. Conversely, the strain stiffening at the inner surface of the active membrane occurs for smaller values of the stretch. The active membrane becomes less compliant and thus its lock-up stretch decreases. Hence, for the active membrane, the change in the stretch associated with the nap-through is smaller. When the passive layer thickness decreases the response of the multilayer tube tends to that of the single-layer one.

\subsection{Electromechanical response for a commercially available dielectric elastomer}

To conclude the numerical analysis, we specialize some of the previous dimensionless results to an active layer made up of a commercially available dielectric elastomer. We consider VHB-4910, a polyacrylate elastomer produced by 3M, that is available as a 1mm thick adhesive. The limiting uniaxial stretch for the VHB can be assumed equal to 7 \citep{PlanteDubowsky2006,KohSuo2}. Typical electromechanical material properties for the VHB-4910 (see, \textit{e.g.}, \citet{Bortot2017JMPS}) are
\beq\lb{VHB4910}
\mu=35\mbox{kPa},\qquad\qquad\epsilon_r=4.5.
\eeq 
The active membrane of the tube, being realized with one layer of VHB-4910, is 1mm thick and we set its radius ratio at 0.8. Thereby, the VHB active membrane is characterized by outer and inner undeformed radii $R_o=5$mm and $R_i=4$mm, respectively. These values correspond to a dimensionless thickness $\widetilde{H}=0.2$.
We consider a single-layer tube, consisting of the VHB active membrane only, and a multilayer one, formed by sandwiching the VHB active membrane between two identical passive layers. The multilayer tube passive layers have the same thickness and shear modulus of the active membrane, $H^{(p)}=H^{(a)}=1$mm and $\mu^{(p)}=\mu^{(a)}=35$kPa.

In the case of axial constraint, the tube electromechanical response is stable (See Fig.~\ref{DEtube_comparison}a). Data comparing the state of the single-layer and the multilayer tubes at 10\% circumferential stretch of the cavity are reported in Tab.~\ref{tab:axconstr}. The voltage required to expand the multilayer tube is approximately one and half time that needed to deform that single-layer one.
\begin{table}[!h]
	\centering
	\small{
		\begin{tabular}{l|l|c|c}\hline
			\multicolumn{4}{c}{\small{Axially constrained tubes}}\\
			\hline
			& &\multicolumn{1}{c|}{\small{neo-Hookean}} &\multicolumn{1}{c}{$\;\;\;$\small{Gent}$\;\;\;$}\\
			\hline
			Single-layer&$\Delta\phi$ [kV] &  15.28  &   15.28  \\
			&$\lambda_{i}$    &   1.10     &    1.10  \\
			\hline
			\hline
			Multilayer&$\Delta\phi$ [kV] &  22.64  & 22.65  \\
			$H^{(p)}=H^{(a)}$ &$\lambda^{(pi)}_{i}$ &  1.10   &   1.10  \\
			$\mu^{(p)}=\mu^{(a)}$ &&&  \\
			\hline
		\end{tabular}}
		\caption{\footnotesize {\textit{Axially constrained tubes}. Comparison, at 10\% circumferential stretch of the cavity, of a single-layer and a multilayer tube, consisting of a 1mm thick VHB-4910 active membrane. The VHB active membrane is characterized by outer and inner radii $R_o=5$mm and $R_i=4$mm, respectively. The multilayer tube has passive layers with the same thickness and the same shear modulus of the active membrane, namely $H^{(p)}=H^{(a)}=1$mm and $\mu^{(p)}=\mu^{(a)}=35$kPa.}}
		\lb{tab:axconstr}
	\end{table}

	In the case of radial constraint, the response of both tube is stable (See Figs.~\ref{DEtube_comparison}b and \ref{DEtube_comparison-z}a). Data comparing the state of the single-layer and the multilayer tubes at 10\% circumferential stretch of the cavity are reported in Tab.~\ref{tab:radconstr}. Whereas the required voltage is similar for the two tubes, the axial deformation achieved by the single-layer tube is much larger than that attained by the multi-layer one.
	\begin{table}[!h]
		\centering
		\small{
			\begin{tabular}{l|l|c|c}\hline
				\multicolumn{4}{c}{\small{Radially constrained case}}\\
				\hline
				& &\multicolumn{1}{c|}{\small{neo-Hookean}} &\multicolumn{1}{c}{$\;\;\;$\small{Gent}$\;\;\;$}\\
				\hline
				Single-layer&$\Delta\phi$ [kV] &  27.11  &   25.72  \\
				&$\lambda_{i}$    &   1.10     &    1.10  \\
				&$\lambda_{z}$    &   1.56     &    1.56  \\
				\hline
				\hline
				Multilayer&$\Delta\phi$ [kV] &  26.87  & 26.88  \\
				$H^{(p)}=H^{(a)}$ &$\lambda^{(pi)}_{i}$ &  1.10   &   1.10  \\
				$\mu^{(p)}=\mu^{(a)}$ &$\lambda_{z}$ &   1.07   &   1.07  \\
				\hline
			\end{tabular}}
			\caption{\footnotesize {\textit{Radially constrained tubes}. Comparison, at 10\% circumferential stretch of the cavity, of a single-layer and a multilayer tube, consisting of a 1mm thick VHB-4910 active membrane. The VHB active membrane is characterized by outer and inner radii $R_o=5$mm and $R_i=4$mm, respectively. The multilayer tube has passive layers with the same thickness and the same shear modulus of the active membrane, namely $H^{(p)}=H^{(a)}=1$mm and $\mu^{(p)}=\mu^{(a)}=35$kPa.}}
			\lb{tab:radconstr}
		\end{table}

		In the unconstrained case, the tubes experience electromechanical instability (See Figs.~\ref{DEtube_comparison}c and ~\ref{DEtube_comparison-z}b). Data comparing the state of the single-layer and the multilayer tubes at the critical state are reported in Tab.~\ref{tab:unconstr}. At the critical state the multilayer tube is more deformed than the single-layer one. Furthermore, the critical voltage of the multilayer tube is almost twice that of the single-layer one.
		\begin{table}[!h]
			\centering
			\small{
				\begin{tabular}{l|l|c|c}\hline
					\multicolumn{4}{c}{\small{Unconstrained tubes}}\\
					\hline
					& &\multicolumn{1}{c|}{\small{neo-Hookean}} &\multicolumn{1}{c}{$\;\;\;$\small{Gent}$\;\;\;$}\\
					\hline
					Single-layer&$\Delta\phi$ [kV] &  20.37  &   20.50  \\
					&$\lambda_{\rmc i}$    &   1.34     &    1.35  \\
					&$\lambda_{\rmc z}$    &   1.26     &    1.27  \\
					\hline
					\hline
					Multilayer&$\Delta\phi$ [kV] &  36.10  & 36.15  \\
					$H^{(p)}=H^{(a)}$ &$\lambda^{(pi)}_{\rmc i}$ &  1.60   &   1.61  \\
					$\mu^{(p)}=\mu^{(a)}$ &$\lambda_{\rmc z}$ &   1.37   &   1.38  \\
					\hline
				\end{tabular}}
				\caption{\footnotesize {\textit{Unconstrained tubes}. Comparison, at the critical state, of a single-layer and a multilayer tube, consisting of a 1mm thick VHB-4910 active membrane. The VHB active membrane is characterized by outer and inner radii $R_o=5$mm and $R_i=4$mm, respectively. The multilayer tube has passive layers with the same thickness and the same shear modulus of the active membrane, namely $H^{(p)}=H^{(a)}=1$mm and $\mu^{(p)}=\mu^{(a)}=35$kPa.}}
				\lb{tab:unconstr}
			\end{table}

			\section{Concluding remarks}\lb{Sec:conclusions}
			
			Electro-active tubes are promising electromechanical transducers, even if their manufacturing is still challenging. In view of practical applications, in the modeling it is necessary to account for insulation of the active membrane to ensure electrode protection, non-perfectly compliant behaviour of the electrodes or interaction of the transducer with a soft actuated body. In order to represent these conditions, a three-layer model can be formulated, in which the active membrane is coated with soft passive layers.
			
			Aiming to understand how the coating layers modify the response of the active membrane, we have investigated the electromechanical response of electro-active tubes formed either by the single active membrane or by a multilayer system, comprising the active membrane and two soft coating layers. Besides the unconstrained case, different constraints have been considered, namely axial and radial constraints.
			
			For pure electric loading, we have shown that the electromechanical response of the multilayer tube is strongly influenced by the shear modulus and by the thickness of the passive layers.
			For \textit{axially constrained} tubes, the electric loading brings about expansion of the tube at fixed axial length. An increase in the shear modulus or in the thickness of the passive layers leads to a proportional increase of the overall stiffness of the multilayer tube. When the passive layers are softer or thinner than the active membrane, the multilayer tube is more compliant and its electromechanical response is closer to that of the single-layer tube. As the passive layers become thicker than the active membrane, this becomes less deformable and its locking stretch decreases.
			For \textit{radially constrained} tubes, the electric loading causes axial extension and expansion of the cavity at fixed outer surface of the tube. The electromechanical response is stable, however when the thickness of the passive layer is twice larger than that of the active membrane electromechanical instability arises. As the passive layers become stiffer, the overall response of the tube becomes stiffer. Anyway, due to the presence of the outer passive layer, the multilayer tube is more deformable with respect to the single-layer one. In particular, the deformation of both the active membrane and the tube cavity increases.
			For \textit{unconstrained} tubes, the electric loading brings about axial extension and expansion of the tube. The electromechanical response is unstable. An increase in the passive layer shear modulus yields an increase in the critical value of $\overline{\Delta \phi}$ and a decrease  in the critical stretch. An increase in the passive layer thickness leads to an increase in both the critical value of $\overline{\Delta \phi}$ and the critical stretch, and to a decrease in the locking stretch of the active membrane. 
			
			Our findings show that the electromechanical response of a multilayer tube can be properly adjusted by modifying the properties of the soft coating layers. This study provide, thereby, new tools for a more reliable modeling of electro-active tube actuators.

			\appendix

			\section{Appendix -- Thin-wall approximation}\lb{sec:Appendix}
			The thin-wall approximation for the case of a single-layer electroactive tube can be obtained by introducing  
			\beq
			\delta= \frac{R_o^2}{R_i^2}-1=\frac{1}{t^2}-1,
			\eeq
			a small parameter tending to zero.
			The stretch at the outer surface of the tube can be expressed as a function of $\delta$
			\beq
			\lambda_o^2=\frac{1}{\delta+1} \left(\lambda_i^2+\frac{\delta}{\lambda_z}\right).
			\eeq
			For an axially constrained neo-Hookean tube, Eq.(\ref{P:axial_constraint_nh}) can hence be rewritten as
			\beq\lb{Thin:P:axial_constraint_nh}
			\begin{split}
			P_i=&\frac{\mu}{2}\left(\frac{\delta(\lambda_i^2\tilde\lambda_z-1)}{\lambda_i^2\tilde\lambda_z(\lambda_i^2\tilde\lambda_z-1+(\delta+1)\tilde\lambda_z^2)}+\frac{2}{\tilde\lambda_z} \ln \left(\sqrt{\frac{\tilde\lambda_z\lambda_i^2(\delta+1)}{\tilde\lambda_z\lambda_i^2+\delta}}\right)\right) \\
			&+ \frac{\epsilon {E_r^0}^2}{2}\left(\sqrt{\frac{1}{\delta+1}}-1
			\right)^2\frac{\delta(\delta+1)}{(\lambda_i^4\tilde\lambda_z+\delta\lambda_i^2) \left[\ln\left(\sqrt{1+\frac{\delta}{\lambda_i^2\tilde\lambda_z}}\right)\right]^2}.
			\end{split}
			\eeq
			For a multilayer tube axially constrained, we can operate in similar way. Each layer is assumed to be thin-walled, so we can introduce 
			\beq
			\delta^{\star}= \frac{1}{{t^{(a)}}^2}-1= \frac{1}{{t^{(pi)}}^2}-1= \frac{1}{{t^{(po)}}^2}-1.
			\eeq
			Assuming that the axial stretch is equal for active and passive layers, the circumferential stretches can thus be expressed as a function of $\delta^{\star}$
			\beq\lb{lambda_relations_thin}
			\begin{split}
				{\lambda_o^{(a)}}^2&=\frac{1}{\lambda_z}+\frac{1}{1+\delta^{\star}}\left({\lambda_i^{(a)}}^2-\frac{\delta^\star}{\lambda_z}\right), \\ {\lambda_i^{(pi)}}^2&=\frac{1}{\lambda_z}+(1+\delta^\star)\left({\lambda_i^{(a)}}^2-\frac{1}{\lambda_z}\right), \\ {\lambda_o^{(po)}}^2&=\frac{1}{\lambda_z}+\frac{1}{(1+\delta^{\star})^2}\left({\lambda_i^{(a)}}^2-\frac{1}{\lambda_z}\right).
			\end{split}
			\eeq
			Substituting these relations in Eq.~(\ref{multi:P:axial_constraint_nh}), we can similarly obtain the thin-wall approximation for the multi-layer tube.

\bibliography{Biblio}

\end{document}